\documentclass[reprint,amsmath,amssymb,aps,pre,tightenlines,bibnotes]{revtex4-2}

\usepackage{graphicx}% Include figure files
\usepackage{dcolumn}% Align table columns on decimal point
\usepackage{bm}% bold math
\usepackage{mathtools}

\usepackage[normalem]{ulem}
\usepackage{color}

\begin{document}
\title{\
Anyonic defect braiding and spontaneous chiral symmetry breaking \\
in dihedral liquid crystals
}
\author{Alexander Mietke}
\email{amietke@mit.edu}

\author{J\"orn~Dunkel}
\email{dunkel@mit.edu}

\affiliation{
Department of Mathematics, Massachusetts Institute of Technology,77 Massachusetts Avenue, Cambridge, Massachusetts 02139-4307, USA
}

\begin{abstract}
Dihedral (\lq $k$-atic\rq) liquid crystals (DLCs) are assemblies of microscopic constituent particles that exhibit $k$-fold discrete rotational and reflection symmetries. Generalizing the half-integer defects in nematic liquid crystals, two-dimensional $k$-atic DLCs can host point defects of fractional topological charge $\pm m/k$. Starting from a generic microscopic model, we derive a unified hydrodynamic description of DLCs with aligning or anti-aligning short-range interactions  in terms of Ginzburg-Landau and Landau-Brazovskii-Swift-Hohenberg theories for a universal complex order-parameter field. Building on this framework, we demonstrate in both particle and continuum simulations how adiabatic braiding protocols, implemented through suitable boundary conditions, can emulate anyonic exchange behavior in a classical system. Analytic solutions and simulations of the mean-field theory further predict a novel spontaneous chiral symmetry breaking transition in anti-aligning DLCs, in quantitative agreement with the patterns observed in particle simulations.
\end{abstract}

\pacs{PACS}

\maketitle
\section{Introduction}
Microscopic particle symmetries fundamentally determine the macroscopic order and dynamics of liquid and crystalline phases of matter~\cite{dege93,chaik00}. Recent technological and experimental progress~\cite{liu16,nied17,fruc20} enables unprecedented precise control over the fabrication and assembly of nanoparticles~\cite{wang03,dong15} and polyhedral colloids~\cite{zhao12,yi13,vutu14,sind14,li16,civan18,loff18} with tunable interactions~\cite{vutu14,wang14,geni18,zhao18}. These advances have led to a renewed theoretical and computational interest in \hbox{$k$-atic} dihedral liquid crystals (DLCs) with discrete $k$-fold rotational and reflection symmetries~\cite{giomi15,ganta15,dussi16,ande17,bowi17,beek17,sart19,mait20,giomi21}. Going beyond the  widely investigated polar ($k=1$) and nematic ($k=2$) liquid crystals~\cite{dege93}, recent studies  showed that assemblies of triatic~\cite{bowi17} ($k=3$) and higher-order polygonal~\cite{ande17} objects ($k\ge 3$) can exhibit striking symmetry breaking phenomena. Furthermore, thanks to seminal work by de Gennes~\cite{1972deGennes}, Halperin and Lubensky~\cite{1974HaLu}, and others ~\cite{1974HaLuMa_PRL,1979PePr,1988Lubensky_PRA,bowi09}, it is well-known that the phenomenological description of 2D liquid crystals shares interesting mathematical similarities~\cite{Zappone17643} with superconductors. Despite their fundamental microscopic differences,  both classes of systems can at the mean-field level be described by a complex field $\Psi(t,\mathbf{r})=|\Psi|e^{i\phi}$ whose magnitude $|\Psi|$ and phase $\phi(t,\mathbf{r})$ encode local order.
\par
A~remarkable characteristic of two-dimensional (2D) DLCs is their ability to host point defects of fractional topological charge~\cite{bowi09,bowi17}, similar to anyonic quasi-particle excitations in 2D quantum matter~\cite{wilc82,lee19}. One of the defining features of anyonic excitations is the behavior of their wave function under pair exchange: When two identical anyons with initial positions $\mathbf{r}_1$ and $\mathbf{r}_2$ are braided counter-clockwise around each other, their complex  wave function $\psi$ changes according to $\psi(\mathbf{r}_2,\mathbf{r}_1)= e^{i2\pi/p}\psi(\mathbf{r}_1,\mathbf{r}_2)$, where $p=1,2,\ldots$. That is, anyonic wave functions acquire a phase $\theta=2\pi/p$ under particle exchange which we refer to as an \textit{anyonic exchange symmetry} in this work; bosons and fermions correspond to the special cases $p=1$ and $p=2$, respectively. First predicted~\cite{lein77} in 1977 and named~\cite{wilc82} in 1982, anyons played an important role in the theoretical explanation of the fractional quantized Hall effect~\cite{laug83,arov84,halp86}. More recently, they have been intensely explored in the context of topological quantum computing~\cite{kita03,chet08}, and  two recent experimental studies~\cite{Nakamura:2019aa,bart20} reported first direct evidence for anyonic quantum statistics. From a general theoretical perspective, the mathematical parallels between the mean-field descriptions of 2D liquid crystals and 2D quantum systems raise the interesting question whether or not one can emulate anyonic exchange symmetries in suitably designed liquid crystal systems.

%%%%%%%%%%%%%%%%%%
%%%%%%%%%%%%%%%%%%
\begin{figure*}[htb]
	\includegraphics[width = 1.95\columnwidth]{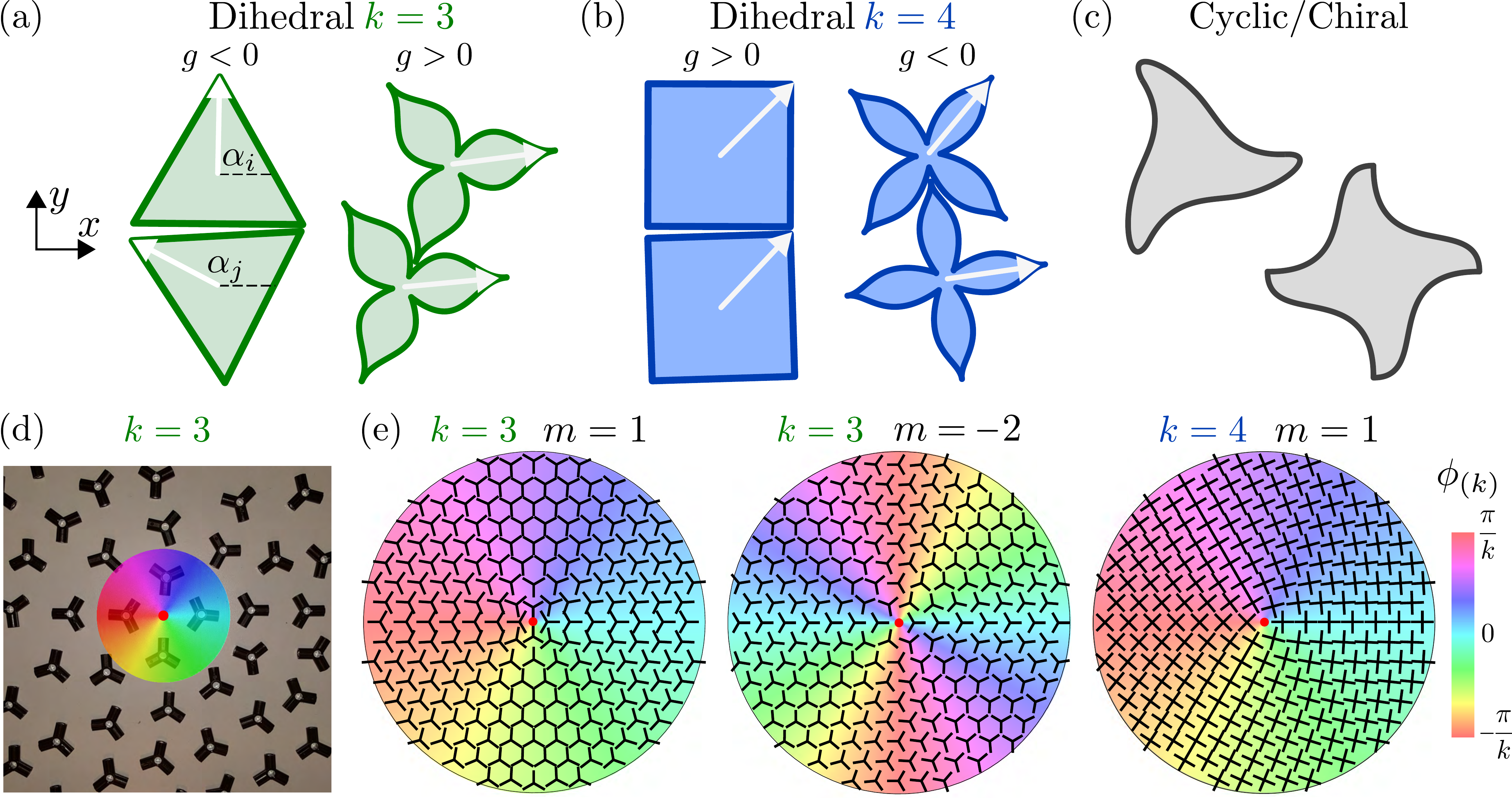}
	\caption{Illustration of DLC particles, $k$-atic order and point defects carrying fractional topological charges $m/k$.
	(a,b)~Examples of 3-atic ($k=3$) and 4-atic ($k=4$) dihedral particles with aligning ($g>0$) and  anti-aligning  ($g<0$) interactions. Director vectors are shown as white arrows.
	(c)~Examples of chiral non-DLC particles that exhibit $k$-fold rotational (cyclic) symmetry but lack reflection symmetry. The present study focuses  exclusively on dihedral particles as in panels (a) and (b). 
	(d)~Realization of  a $+1/3$-defect in collection of 3-legged symmetric LEGO connectors with mutually repulsive magnetic interactions. For additional experimental realizations of $k$-atic liquid cystals in colloidal suspensions see Fig.~1 in~\cite{giomi21}.
	(e)~From left to right: $k$-atic phase angle fields $\phi_{(k)}$ for point defects of topological charge $+1/3$  in a 3-atic ($k=3,m=1$), charge $-2/3$ in a 3-atic  ($k=3,m=-2$) and  charge $+1/4$  in a 4-atic field ($k=4,m=1$) with local $k$-atic directors overlaid (see also Fig.~\ref{figS0} for additional examples). 
	}
	\label{fig1}
\end{figure*}
%%%%%%%%%%%%%%%%%%
%%%%%%%%%%%%%%%%%%

Here, we will show that it is indeed possible to create, stabilize and manipulate pairs of identical fractional defects $(\mathbf{r}_1,\mathbf{r}_2)$ in $k$-atic DLCs, such that a braiding exchange $(\mathbf{r}_1,\mathbf{r}_2)\rightarrow(\mathbf{r}_2,\mathbf{r}_1)$ is accompanied by the accumulation of a globally constant $k$-atic phase difference in the order parameter field $\Psi_k$. To this end, we first derive a hydrodynamic mean-field description of DLCs by starting from a generic $XY$-type particle interaction model on a random lattice. We then verify that the derived mean-field theory agrees quantitatively with simulations of the particle model. After this validation step, we demonstrate anyon-like exchange symmetries in the particle model by using adiabatically modulated boundary anchoring conditions to braid a pair of topological defects. Finally, we identify a novel chiral symmetry-breaking transition in classical DLCs with anti-aligning short-range interactions, which manifests itself through the spontaneous formation of stable braidable spiral patterns in the phase angle fields. The close agreement between analytical and numerical solutions of the mean-field theory and particle simulations suggests that the theoretical predictions presented below could be realized with a variety of experimental systems. Candidates range from 3-fold symmetric molecules (e.g. 1,3,5-trichlorobenzene~as proposed in \cite{bowi17}) or DNA-origami structures~\cite{Chao:2018aa,Veneziano1534,Siavashpouri:2017aa} to $k$-atic colloidal platelets~\cite{zhao12,loff18} and polyhedral~\cite{nied17},~magnetic~\cite{soni19} or DNA-coated colloids~\cite{C7SM01722G,yi13} (see also Fig.~1 in Ref.~\cite{giomi21}).
%\newpage

%%%%%%%%%%%%%%%%%%%%%%
%%%%%%%%%%%%%%%%%%%%%%
\section{Mean-field description of 2D DLC\lowercase{s} with aligning and anti-aligning interactions}
%%%%%%%%%%%%%%%%%%%%%%
%%%%%%%%%%%%%%%%%%%%%%

Generalizations of polar and nematic liquid crystals to higher-order symmetry groups are often called $k$-atics~\cite{bowi09}, where the integer $k$ counts the discrete rotational symmetries of the constituent particles (Fig.~\ref{fig1}a,b). Here, we focus on systems of particles that have both $k$-fold discrete rotational and reflection symmetry (Fig.~\ref{fig1}a,b). This set of symmetry transformation defines the dihedral group $D_k$, which maps a regular polygon with $k$ corners onto itself. Accordingly, we will refer to such systems generically as dihedral liquid crystals (DLCs).  Note that the invariance  under reflections excludes chiral particles, which can still have discrete rotational (cyclic) symmetry (Fig.~\ref{fig1}c). Depending on whether $k$ is even or odd, and whether the effective particle shape is convex or concave, DLCs can have aligning or anti-aligning interactions (Fig.~\ref{fig1}a,b). Here, \lq shape\rq{} can be understood in a broader sense as the angular symmetry of the particle's pair-interaction potential, for which we will provide a concrete example in Sec.~\ref{sec:cg}. As we will show in detail below, monodisperse DLCs with aligning and anti-aligning interactions can be described by a single universal mean-field equation. Before delving into the more technical discussion, it is instructive to anticipate the structure of the resulting mean-field equations and their relations to anyonic exchange symmetries.

%%%%%%%%%%%%%%%%%%%%%%
\subsection{Universal mean-field equation}
%%%%%%%%%%%%%%%%%%%%%%

In the limit of a constant particle density, a unified mean-field description of 2D monodisperse DLCs can be given in terms of a complex-valued scalar order parameter field $\Psi_k(t,\mathbf{r})$ that is governed by the relaxation dynamics
\begin{align}\hspace{-0.1cm}
\tau\partial_t\Psi_k&=-\left(A+B|\Psi_k|^2\right)\Psi_k+\mathcal{L}\left(\nabla^2\right)\Psi_k=-\frac{\delta\mathcal{E}_k}{\delta\Psi_k^*}.
\label{eq:RGLgen}
\end{align}
Here, $\mathcal{L}\left(\nabla^2\right)$ denotes a linear differential operator, and $\mathcal{E}_k$ is the corresponding energy functional. The subscript $k$ indicates the $k$-fold symmetry, and~$\tau$ is a relaxation time-scale that can be computed from the microscopic particle dynamics. The magnitude~$|\Psi_k|$ characterizes the degree of local order (alignment) of the director unit vectors (white arrows in Fig.~\ref{fig1}a,b), and the $k$-atic phase angle of $\Psi_k$ their mean orientation. 
\par
As shown in detail below, the real parameters $A$ and~$B$ depend on the particle symmetry and interaction strength. For aligning interactions, the operator~$\mathcal{L}$ in Eq.~(\ref{eq:RGLgen}) reduces to a Laplacian  $\mathcal{L}=L^2\nabla^2$; in this case, Eq.~\eqref{eq:RGLgen}  corresponds a \lq real\rq\ Ginzburg-Landau (GL) equation~\cite{aran02} with an effective diffusion constant $D=L^2/\tau$ (Sec.~\ref{sec:any}). 
For anti-aligning interactions, $\mathcal{L}$ will take the form of a pattern-forming Swift-Hohenberg-type (SH)~\cite{cross09} operator \mbox{$\mathcal{L}=-L_1^2 \nabla^2 - L_2^4 (\nabla^2)^2$} (Sec.~\ref{sec:SH}). 
\par
Conceptually, Eq.~\eqref{eq:RGLgen} formalizes the mathematical correspondence between the mean-field descriptions of aligning DLCs and quantum fluids. If  $\mathcal{L}$ is proportional to the Laplacian, the energy $\mathcal{E}_k$ in Eq.~(\ref{eq:RGLgen}) takes the form~\cite{bowi09}~(Appendix~\ref{app:freeEnLCs})
\begin{equation}\label{eq:ComplexGLE}
\mathcal{E}_k=\int d^2r\left(A\left|\Psi_k\right|^2+\frac{B}{2}\left|\Psi_k\right|^4+L^2\left|\nabla\Psi_k\right|^2\right).
\end{equation}
Then, for $k=1$ and $\tau=i$ in Eq.~\eqref{eq:RGLgen}, one recovers the Gross-Pitaevskii equation~\cite{gross61,hein19} describing Bose-Einstein condensates. Thus, 2D aligning DLCs and quantum fluids can be considered energetically equivalent at the mean-field level, while differing by the fact that the former have dissipative dynamics  whereas the latter have conservative dynamics.  Similar mean-field correspondences played a historically important role for the understanding of smectic liquid crystal phases by their analogy with superconductors~\cite{1972deGennes,1974HaLu,1988Lubensky_PRA}. 

For the SH-type mean-field theory of DLCs with anti-aligning interactions, $\mathcal{E}_k$ in Eq.~\eqref{eq:RGLgen} corresponds to a Landau-Brazovskii (LB) energy~\cite{braz75} (Appendix~\ref{app:freeEnSH}). The LB energy functional generally captures the mean-field dynamics of systems with competing microscopic interactions, such as diblock copolymeres~\cite{bate90,spen13} or microemulsions~\cite{ciach13,care20}, and it played a key role in explaining how fluctuations affect the properties of order-disorder phase transitions~\cite{braz75,swift77,hohen77,jano13}.

In the context of our present study, Eq.~\eqref{eq:RGLgen} provides the basis for realizing classical counterparts of anyon exchange symmetries. More specifically, we will see below that Eq.~(\ref{eq:RGLgen}) accurately describes the formation, stability and decay (Figs.~\ref{fig2}, \ref{fig4}), and the braiding (Figs.~\ref{fig3}, \ref{fig5}) of fractional topological defects as observed in particle simulations of generic $XY$-type microscopic DLC models. When appropriately braided around each other, these fractional topological defects mimic the behavior of anyonic quasi-particle excitations by acquiring a phase-shift in the complex order parameter field.

%%%%%%%%%%%%%%%%%%%%%%
\subsection{Fractional topological charges}
%%%%%%%%%%%%%%%%%%%%%%

To characterize the orientational order of DLCs, one can express the complex order-parameter field in the polar form~(Appendix~\ref{app:freeEnLCs})
\begin{equation}\label{eq:psik}
\Psi_k=|\Psi_k| e^{ik\phi_{(k)}}
\end{equation}
The magnitude field $|\Psi_k|$ measures the strength of the local $k$-atic order, and the phase angle field  
\begin{equation}\label{eq:phasedef}
\phi_{(k)}=\frac{\arg\Psi_k}{k}
\end{equation}
indicates the mean $k$-atic director orientation. The widely studied polar and nematic liquid crystals correspond to  $k=1$ and $k=2$,  with microscopic constituents  symmetric under rotations of $2\pi$ or $\pi$, respectively.
\par
The phase field~$\phi_{(k)}$ of a $k$-fold symmetric DLC can host commensurate fractional point defects~(Appendix~\ref{appsec:ptdef}). The net topological defect charge $q_\text{d}$ enclosed by a positively oriented curve $\mathcal{C}$ is obtained as
\begin{equation}\label{eq:defch}
q_\text{d}=\frac{1}{2\pi}\oint_{\mathcal{C}}d\mathbf{l}\cdot\nabla\phi_{(k)}=\frac{m}{k},
\end{equation}
where $m$ can be any integer. Examples of defect states in $k$-atic DLCs are illustrated in Figs.~\ref{fig1}d,e and~\ref{figS0}. The \lq experimental\rq{} realization of +1/3-defect in Fig.~\ref{fig1}d was assembled from 3-atic LEGO  toy elements  that are invariant under $2\pi/3$-rotations and corresponding reflections, and carry repulsive magnetic dipoles in each of their legs. Owing to  the discrete symmetry of each microscopic element, only a rotation by $2\pi/k$ is required before a particle returns to its initial configuration. If particle orientations along the integration contour complete $|m|$ counterclockwise rotations in total then the defect charge is positive with $m>0$ in Eq.~(\ref{eq:defch}), whereas $m<0$ signals the completion of $|m|$ clockwise $2\pi/k$-rotations along the~curve~$\mathcal{C}$.

%%%%%%%%%%%%%%%%%%%%%%
%%%%%%%%%%%%%%%%%%%%%%
\section{Microscopic DLC model}
\label{sec:cg}
%%%%%%%%%%%%%%%%%%%%%%
%%%%%%%%%%%%%%%%%%%%%%

\textit{A priori}, it is not clear how well a mean-field model can capture the behavior of a specific microscopic DLC system. To validate predictions obtained from Eq.~\eqref{eq:RGLgen}, we will compare them against simulations of a generic microscopic DLC model that can realize both aligning and anti-aligning interactions. Specifically, we consider an $XY$-type model~\cite{kost73} describing particles with local orientation angles~\hbox{$\alpha_i(t)\in[0,2\pi)$} that interact according to the overdamped dynamics
\begin{equation}
\frac{d\alpha_i}{dt}=\frac{g}{\pi R_{\alpha}^2}\sum_{j\in\mathcal{N}_i}\sin\left[k(\alpha_j-\alpha_i)\right]+\sqrt{2D_r}\xi_i(t).
\label{eq:micrmod}
\end{equation}
Equation~\eqref{eq:micrmod} is invariant under rotations $\alpha_i\rightarrow\alpha_i+2\pi/k$ and corresponding reflections along the symmetry axis of the particles.  The parameter $g$ sets the effective interaction strength between particle $i$ and particles $j$ in a neighborhood~$\mathcal{N}_i$ of radius $R_{\alpha}$. For $g>0$, the $k$-atic directors of nearby particles tend to align, whereas $g<0$ favors anti-alignment (Fig.~1a,b). The Gaussian white noise~$\xi_i(t)$ has zero mean, satisfies \hbox{$\langle\xi_i(t)\xi_j(t')\rangle=\delta_{ij}\delta(t-t')$}, and $D_r$ is the rotational diffusion constant. In all simulations presented below, particles were randomly placed inside a circular disk domain and given time to redistribute homogeneously through isotropic short-range repulsion~(Appendix~\ref{appsec:partsim}). Thereafter, the particle positions were held fixed and the angular dynamics Eq.~\eqref{eq:micrmod} was turned on. This model may thus be interpreted as a generalized classical $XY$-model on a densely packed random lattice~\cite{mcar86}.

%%%%%%%%%%%%%%%%%%%%%%
\subsection{Discrete $k$-scaling invariance}
%%%%%%%%%%%%%%%%%%%%%%

\begin{figure*}
\centering
	\includegraphics[width = 1.95\columnwidth]{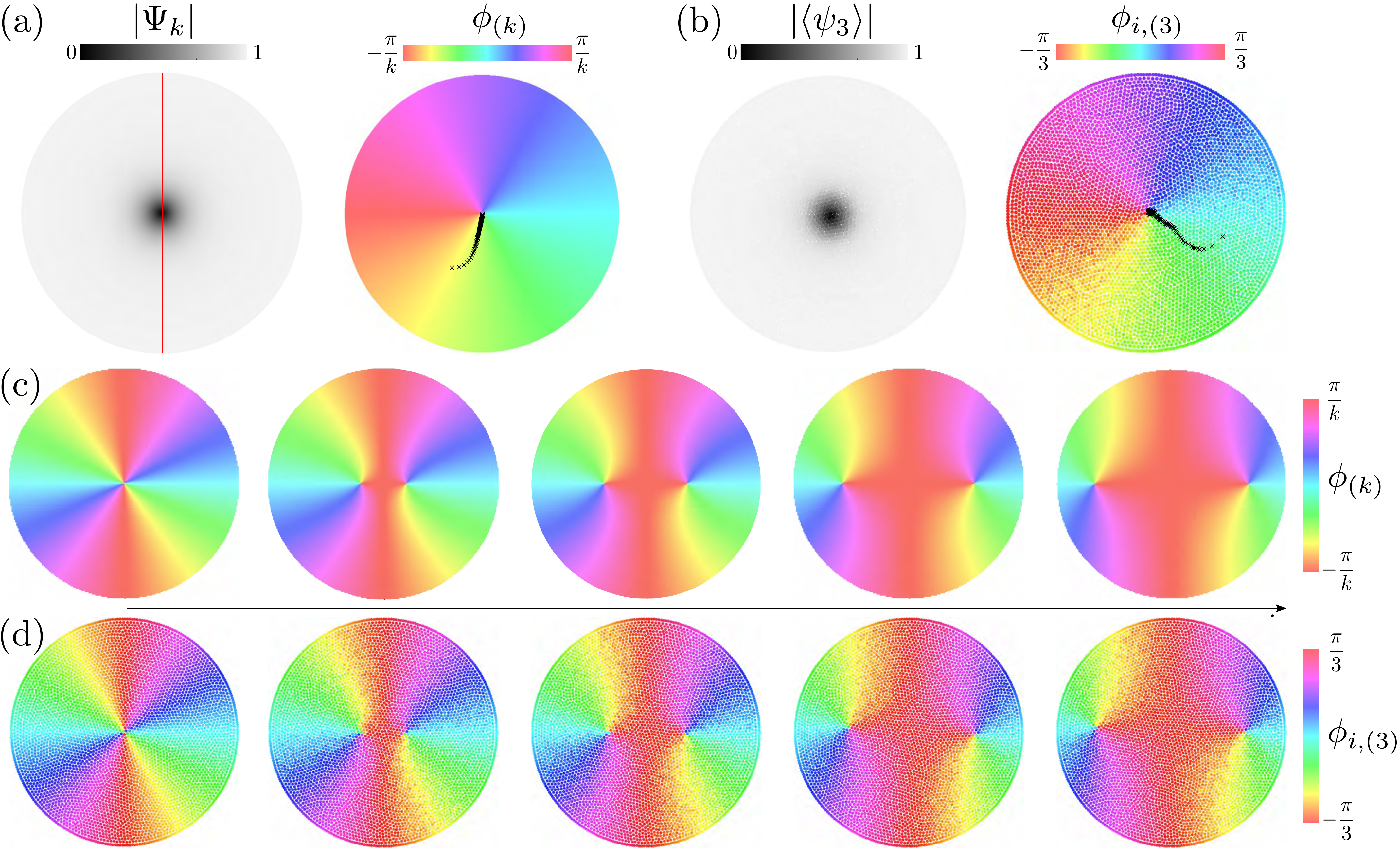}
	\caption{Mean-field theory correctly predicts defect formation, stability and decay as seen in the particle model with aligning interactions ($g>0$) on a unit disk.
	(a)~Stationary $1/k$-defect solution emerging from random initial conditions in the mean-field Eq.~\eqref{eq:RGLgen} with $\mathcal{L}=L^2\nabla^2$, $A=-1$, $B=1$, $L=0.08$, for boundary anchoring Eq.~\eqref{eq:bc1} with $m=1$ and $\gamma_k=\theta$. The lines $\text{Re}\Psi_k=0$ (red) and $\text{Im}\Psi_k=0$ (blue) indicate how the boundary anchoring induces a point defect with $|\Psi_k|=0$ inside the unit disk. The defect trajectory (black) shows that the defect appears at a random position in the bulk and subsequently moves towards the center of the disk (Movie~1).
	(b)~Steady-state $1/3$-defect emerging from random initial conditions in particle simulations of Eq.~(\ref{eq:micrmod}) with $k=3$, using an equivalent orientational boundary anchoring Eq.~(\ref{eq:bcpm}) with $\gamma_k=\theta$. $\langle \psi_3\rangle$ denotes the time-average of the local $3$-atic order parameter \smash{$\psi_3=\sum_{j\in\mathcal{N}_i}\exp\left(3i\alpha_j\right)/|\mathcal{N}_i|$} at steady state, and~$\phi_{i,(k)}=\text{arg}\left(e^{ik\alpha_i}\right)/k$ is the instantaneous $k$-atic director orientation. In agreement with  the mean-field prediction,  the defect trajectory (black) shows that a phase defect forms at a random position in the bulk and moves stochastically towards the center of the disk (Movie~1). 
	(c,d)~For profiles $\gamma_k=\theta$ and $m=2$ in boundary anchorings Eqs.~(\ref{eq:bc1}) and (\ref{eq:bcpm}), a $2/k$-defect decays into two $1/k$-defects in both the mean-field simulations (c) and 3-atic particle simulations~(d) (Movie~2). Snapshots show instantaneous phase fields at $t/\tau\in\{0,7.5,15,37.5,150\}$ [Eq.~\eqref{eq:RGLgen}; c] and 
$t/\bar{\tau}\in\{0,18,36,80,320\}$ (particle model; d). Simulations of~Eq.~\eqref{eq:micrmod} used 4,000 non-anchored bulk particles, 900 anchored boundary particles, and parameters $D_r=1$ (rotational diffusion chosen as characteristic time scale of particle simulations), $R_{\alpha}=0.2$ (interaction radius) and $g=0.25$ (interaction strength).	
	}
	\label{fig2}
\end{figure*}

In agreement with mean-field predictions, simulations for particles with $k$-fold symmetry show topological defects of fractional charges that are integer multiples of~$1/k$ (Fig.~\ref{fig2}). We note, however, that for any value of $k$, the dynamics given in Eq.~\eqref{eq:micrmod} can be mapped onto an equivalent polar model with $k=1$, by defining  rescaled director angles $\alpha_i'=k\alpha_i$, a rescaled alignment strength $g'=gk$ and a rescaled rotational diffusion \hbox{$D'_r=D_rk^2$}. The fact that Eq.~\eqref{eq:micrmod} can be rescaled in this form essentially explains why the DLC particle models with different~$k$ can be described by the same mean-field Eq.~\eqref{eq:RGLgen}. Although, as we will discuss next, the coefficients in Eq.~\eqref{eq:RGLgen} depend on~$k$, the structure of the mean-field equation remains preserved for particles with different dihedral symmetries.

%%%%%%%%%%%%%%%%%%%%%%
\subsection{Mean-field parameters}
%%%%%%%%%%%%%%%%%%%%%%
To show how Eq.~\eqref{eq:RGLgen} can be derived from the microscopic model in Eq.~\eqref{eq:micrmod}, we generalize standard coarse-graining procedures~\cite{dean96,bert09,farr12} to the case of $k$-atic particle interactions with a finite spatial range~\hbox{\cite{gros14,ken15,arol20}}~(Appendix~\ref{app:CG}). To this end, we decompose the one-particle probability density function of the $N$-particle system, which is defined by the Gaussian white-noise average
\begin{equation}
f(\alpha,\mathbf{r},t)=\sum_{i=1}^{N}\langle\delta\left(\alpha-\alpha_i(t)\right)\delta\left(\mathbf{r}-\mathbf{r}_i\right)\rangle,
\label{eq:fprobdef}
\end{equation}
into its angular moments
\begin{equation}
f_{n}(\mathbf{r},t)=\int_0^{2\pi} d\alpha\, f(\alpha,\mathbf{r},t)\,e^{in\alpha}.
\label{eq:Fourierang}
\end{equation} 
The mode $f_0$ represents the particle number density $\rho$ of the system, which in our case is homogeneous and fixed. Accordingly, we define normalized and dimensionless modes by
\begin{equation}
\psi_n=\frac{f_n}{\rho}.
\end{equation}
Equations~\eqref{eq:micrmod} and~\eqref{eq:Fourierang} yield an infinite hierarchy of dynamic equations for the complex modes~$\psi_n$~(Appendix~\ref{app:CG}). Despite being nonlinear, these equations decouple modes with $n=jk$ for integers $j\ne0$ from all modes with $n\ne jk$, which can be understood as a consequence of the $k$-rescaling property of the microscopic model in~Eq.~(\ref{eq:micrmod}). In the limit of a vanishing interaction radius $R_{\alpha}\rightarrow0$, one then finds the spatially homogeneous dynamics (Appendix~\ref{app:CG-1})
\begin{equation}
\bar{\tau}\partial_t\psi_k=-\left(\bar{A}+\bar{B}|\psi_k|^2\right)\psi_k=:h_k,
\label{eq:kcoarse}
\end{equation}
with characteristic relaxation time-scale $\bar{\tau}=2/(|g|k\rho)$ and parameters 
\begin{equation}
\bar{A}=-\text{sgn}(g)\left(1-\frac{2D_rk}{g\rho}\right),
\label{eq:abar}
\qquad
\bar{B}=\frac{|g|\rho}{4D_rk}.
\end{equation}
By comparing Eqs.~\eqref{eq:kcoarse} and \eqref{eq:abar} with Eq.~\eqref{eq:RGLgen}, we can identify $\psi_k\simeq\Psi_k$, $\bar{\tau}\simeq\tau$, $\bar{A}\simeq A$ and $\bar{B}\simeq B$, indicating that the particle dynamics~\eqref{eq:micrmod} indeed provides a microscopic realization of the mean-field theory~\eqref{eq:RGLgen}.  Below, we extent Eq.~\eqref{eq:kcoarse} to short-range interactions with $R_{\alpha}>0$, which leads to leading-order corrections in the form of linear operators~$\mathcal{L}(\nabla^2)$ as indicated in~Eq.~\eqref{eq:RGLgen}.
\par
Beforehand, we note that Eq.~(\ref{eq:kcoarse}) generalizes the corresponding result~\cite{bert09} for polar systems (\hbox{$k=1$}) to arbitrary $k$-atic systems. A change in the sign of the  coefficient $\bar{A}$ signals the spontaneous emergence of homogeneous $k$-atic order due to a linear instability of Eq.~(\ref{eq:kcoarse}) if \hbox{$g>g^*:={2kD_r}/{\rho}>0$}, corresponding to an instability at wave vector~$\mathbf{q}=0$. This also implies that for anti-aligning interactions with $g<0$, the disordered state is linearly stable in the limit of point-wise interactions. 
\par
In the remainder, we will consider the experimentally relevant  case of systems with finite interaction range $R_{\alpha}>0$.  By comparing the mean-field predictions  of Eq.~(\ref{eq:RGLgen}) with quantitatively mapped microscopic models described by Eq.~(\ref{eq:micrmod}), we will demonstrate the controlled manipulation of fractional defects through boundary anchoring (Sec.~\ref{sec:any}) and the spontaneous formation of chiral textures (Sec.~\ref{sec:SH}).

%%%%%%%%%%%%%%%%%%
%%%%%%%%%%%%%%%%%%
%%%%%%%%%%%%%%%%%%
%%%%%%%%%%%%%%%%%%

%%%%%%%%%%%%%%%%%%%%%%
%%%%%%%%%%%%%%%%%%%%%%
\section{Defect braiding and anyonic exchange symmetries in aligning DLC\lowercase{s}~($g>0$)}
\label{sec:any}
%%%%%%%%%%%%%%%%%%%%%%
%%%%%%%%%%%%%%%%%%%%%%

We first show that for aligning short-range interactions in the particle model Eq.~(\ref{eq:micrmod}), the mean-field description~(\ref{eq:RGLgen}) takes the form of a \lq real\rq\ GL equation. Thereafter, we identify boundary anchoring conditions that will allow us to position and manipulate fractional defects in both mean-field and particle  simulations. We will then apply this framework to braid two identical point defects, which gives rise to an emergent anyonic exchange symmetry in the particle model. Finally, a protocol of boundary anchoring modulations is proposed for which a global $k$-atic phase change arises in complex order parameter fields described by both mean-field and particle simulations.

%%%%%%%%%%%%%%%%%%%%%%%%%%
\subsection{GL mean-field theory for aligning $k$-atics}
%%%%%%%%%%%%%%%%%%%%%%%%%%

Assuming aligning interactions \mbox{($g>0$; Fig.~\ref{fig1}a,b)} and an isotropic interaction neighborhood in the particle model Eq.~(\ref{eq:micrmod}), the spatio-temporal $k$-atic mode dynamics can be approximated by~(Appendix~\ref{appsec:psdiffop})
\begin{equation}
\bar{\tau}\partial_t\psi_k(\mathbf{r},t)\approx h_k+\frac{R_{\alpha}^2}{8}\nabla^2\psi_k(\mathbf{r},t),
\label{eq:kcoarseany}
\end{equation} 
with the homogeneous part $h_k$ given in Eq.~(\ref{eq:kcoarse}). Note that the effective diffusion constant \hbox{$\bar{D}=R_{\alpha}^2/(8\bar{\tau})$} is not the result of actual particle diffusion but instead arises  from the finite-range interactions between particles at fixed positions.
\par
Equation~(\ref{eq:kcoarseany}) shows that the mean-field description of aligning DLC particle systems is given by Eq.~(\ref{eq:RGLgen}) with
\begin{equation}\label{eq:Ldef}
\mathcal{L}=L^2\nabla^2,
\qquad 
L\simeq\frac{R_{\alpha}}{\sqrt{8}},
\end{equation}
corresponding to a \lq real\rq\ GL equation~\cite{aran02} for the $k$-atic order parameter $\Psi_k$. The correlation length $L$ is related to an effective bending rigidity  that penalizes deviations of the $k$-atic  director field from a homogeneously aligned state~\cite{chand92,dege93}~(Appendix~\ref{app:freeEnLCs}). We will now exploit  this feature to generate and position topological defects through appropriate boundary conditions.

%%%%%%%%%%%%%%%%%%%%%%%%%%
\subsection{Defect positioning through boundary anchoring}
%%%%%%%%%%%%%%%%%%%%%%%%%%

To illustrate how the total topological charge and the positioning of $k$-atic defects can be controlled, we consider numerical solutions of the GL equation~(\ref{eq:RGLgen}) on a unit disk domain $S$ with boundary anchoring
\begin{equation}\label{eq:bc1}
\Psi_k|_{\partial S}=e^{im\gamma_k}.
\end{equation}
Through a prescribed \textit{anchoring profile} $\gamma_k(\theta)$, Eq.~(\ref{eq:bc1}) fixes the orientation of the $k$-atic director along the boundary~$\partial S$ to
\begin{equation}
\phi_{(k)}=\frac{1}{k}\text{arg}\left(e^{im\gamma_k}\right).
\label{eq:anglemap}
\end{equation} 
The examples discussed in the remainder are based on a monotonically increasing profile $\gamma_k(\theta)$ with \hbox{$\gamma_k(2\pi)-\gamma_k(0)=2\pi$}, for which the boundary condition~(\ref{eq:bc1}) imposes a total topological charge $m/k$ in the disk. 

The simplest nontrivial anchoring condition~\eqref{eq:bc1}, corresponding to a topological net charge of~$1/k$ in the disk, is $m=1$ and $\gamma_k=\theta$. In this case, the GL relaxation dynamics  favors the formation of a stationary $1/k$-defect at the center of the disk~(Fig.~\ref{fig2}a; Movie~1). Although this might have been expected on symmetry grounds, it is effectively a consequence of the director field's bending rigidity mediated by $L$: A central position reduces the distortions of the director field around the defect to a single, monotonous winding that is minimally necessary to be compatible with the boundary conditions. 

%%%%%%%%%%%%%%%%%%
%%%%%%%%%%%%%%%%%%
\begin{figure*}%[t]
\centering
	\includegraphics[width = 1.98\columnwidth]{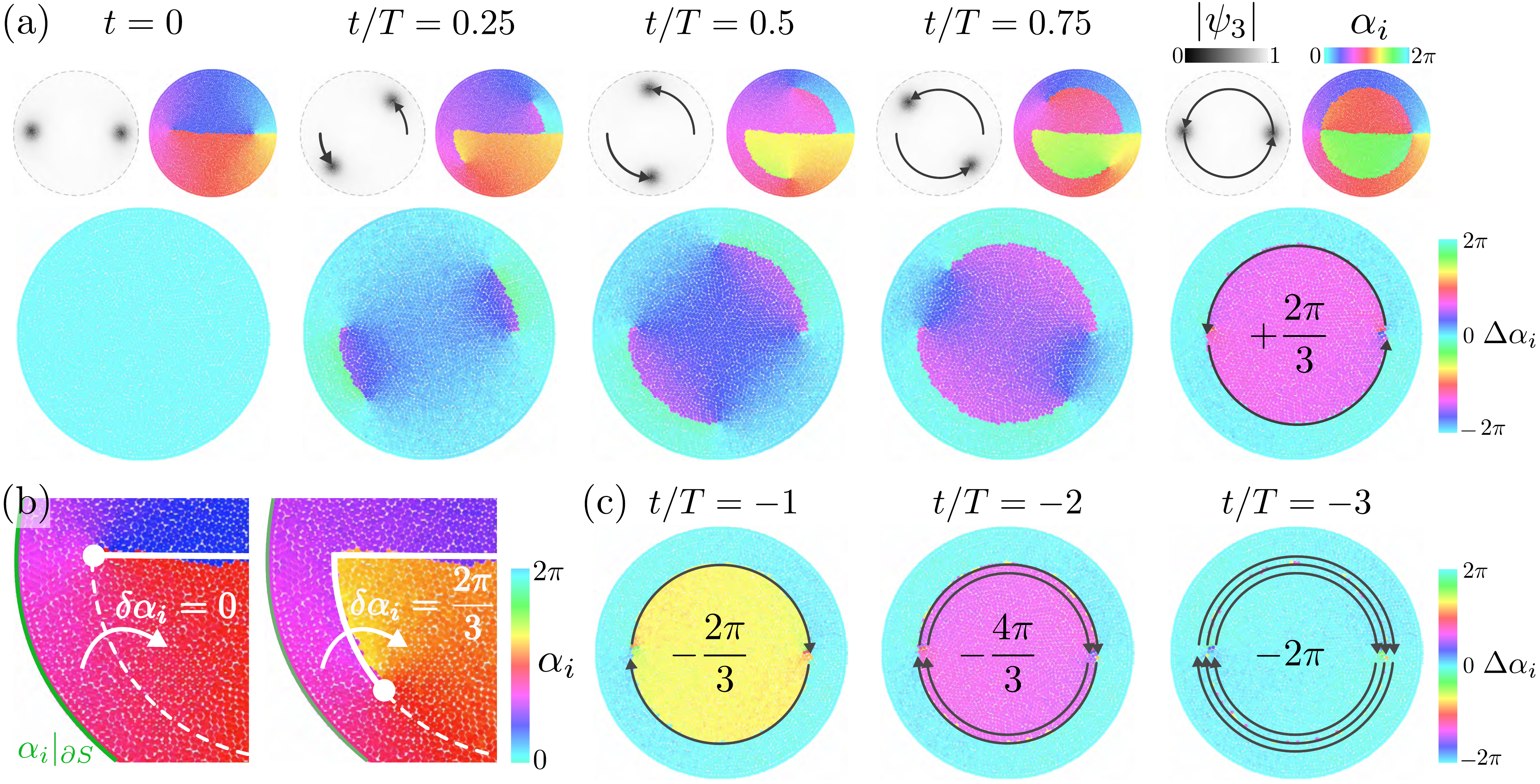}
	\caption{Defect braiding in $k$-atic particle simulations with aligning interactions ($g>0$) gives rise to emergent anyonic exchange symmetries. (a)~A pair of symmetric $1/3$-defects in the 3-atic particle model on a unit disk (the final state of Fig.~\ref{fig2}d) is braided \textit{counter-clockwise} through a modulation of the boundary anchoring profile (Eqs.~\eqref{eq:bcpm}--\eqref{eq:t0param} with $\theta_0=0$ and $t/T\in[0,1]$). $\psi_3$~denotes the local $3$-atic order parameter [see Fig.~\ref{fig2}b and Eq.~(\ref{eq:psikdef})], and $\Delta\alpha_i:=\alpha_i(t)-\alpha_i(t=0)$ measures the change of a particle's orientation angle. After one defect exchange ($t/T=1$, rightmost images) the domain enclosed by the defect trajectories (black arrows) picks up a globally uniform orientation angle difference $\Delta\alpha_i=2\pi/3$. 
	(b)~Enlarged portion of particle angles $\alpha_i$ shown in (a) (left: $t/T=0$, right: $t/T=0.25$) reveal the mechanism by which the inner domain picks up a well-defined angle difference: The orientation of particles near the boundary is held fixed through the boundary anchoring~$\alpha_i|_{\partial S}$. Initially (left), particle angles change continuously ($\delta\alpha_i=0$) along the radial direction. The $1/3$-defects (white dot) are connected by a branch cut (white solid line) across which particle orientation angles $\alpha_i$ jump by $2\pi/3$. Defect braiding represents a continuous deformation of the branch cut (right), which introduces a discontinuous jump $\delta\alpha_i=2\pi/3$ relative to (fixed) particle orientations near the disk boundary. (c)~Running the braiding protocol backwards leads to \textit{clockwise} braiding and opposite signs of $\Delta\alpha_i$. For the case $k=3$ illustrated here, the system returns to its initial state after three successive braiding steps, in analogy  with anyonic excitations in quantum systems (Movie~3). All simulations use the boundary anchoring given in Eqs.~(\ref{eq:anchwind}) and (\ref{eq:t0param}) with $a=0.1$ and $D_rT=20$; all other parameters are as in~Fig.~\ref{fig2}d.
	}
	\label{fig2x}
\end{figure*}
%%%%%%%%%%%%%%%%%%
%%%%%%%%%%%%%%%%%%

When imposing $m=2$ in Eq.~\eqref{eq:bc1}, the net topological charge in the disk is~$2/k$. For suitable initial conditions, this charge may first be concentrated in a single defect, which then splits into a pair of $1/k$-defects with lower energy (Fig.~\ref{fig2}c; Movie~2). The final steady state reflects that it is energetically favorable to distribute the director field distortions around two $1/k$ defects, instead of winding the director field symmetrically but with twice the rate around a single $2/k$ defect. Similarly for $m>2$ and $\gamma_k=\theta$, transient higher-charge defects decay into $m$ single $1/k$-defects that eventually settle into symmetric low-energy configurations~(Fig.~\ref{figS2}a,b).

To test if these predictions can indeed be reproduced in the particle model, we simulated Eqs.~\eqref{eq:micrmod} on a unit disk (Fig.~\ref{fig2}b,d), where a particle orientation~$\alpha_i$ corresponds to a $k$-atic director orientation 
\begin{equation}
\phi_{i,(k)}=\frac{1}{k}\text{arg}\left(e^{ik\alpha_i}\right).
\label{eq:micrangmap}
\end{equation}
Comparing this with Eq.~(\ref{eq:anglemap}), the mean-field boundary condition~\eqref{eq:bc1} can be matched by fixing the orientations of particles at the boundary to
\begin{equation}\label{eq:bcpm}
\alpha_i|_{\partial S}=({m}/{k})\gamma_k.
\end{equation}
Note that particle orientations $\alpha_i$ and $k$-atic director orientations $\phi_{i,(k)}$ [Eq.~(\ref{eq:micrmod})] are two distinct observables. In particular, $\phi_{i,(k)}\in(-\pi/k,\pi/k]$ can be unambiguously determined for dihedral shapes with $k$ sides, while measuring a corresponding particle orientation $\alpha_i\in[0,2\pi)$ requires an additional polar feature as indicated by the white arrows in Fig.~\ref{fig1}. For example, in $k$-atic colloidal systems~\cite{zhao12,loff18} it would suffice to mark one corner of each platelet for tracking purposes.
\par
Using the same monotonous boundary anchoring \mbox{$\gamma_k=\theta$}, we find that the resulting particle simulations~(Fig.~\ref{fig2}b,d and Fig.~\ref{figS2}a,b) agree well with the GL theory. For example, for $k=3$ and $m=1$, a single $1/3$-defect forms and  moves to the center of the domain (Fig.~\ref{fig2}b; Movie~1), whereas for $k=3$ and $m=2$, an initially created $2/3$-defect splits into two $1/3$-defects that move symmetrically away from each other until they reach a symmetric steady state position  (Fig.~\ref{fig2}d; Movie~2). In both cases, the final steady state textures confirm the GL mean-field prediction.

The above examples illustrate how topological defects take equilibrium positions that effectively minimize director winding gradients around them. It follows that for a constant azimuthal anchor-winding gradient ($\partial_{\theta}\gamma_k=$ const.) at the disk boundary, the equilibrium positions of defect-pairs are degenerate with respect to rotations around the disk center. In turn, structured anchoring enables a targeted defect positioning. We demonstrate this useful fact by controlling the orientation of the axis connecting a $1/k$-defect pair. To this end, we consider Eq.~\eqref{eq:bcpm} with $m=2$ and anchoring profile
\begin{equation}
\label{eq:anchwind}
\gamma_k(\theta)=\theta-\theta_0+a\sin\left[2(\theta-\theta_{\textrm{a}})\right],
\end{equation}
where $a$ sets the strength of the defect anchoring and $|a|<1/2$ ensures that $\gamma_k(\theta)$ is monotonic. This choice of $\gamma_k(\theta)$ motivated as follows (consider $\theta_0=0$ for simplicity): For $\theta_{\textrm{a}}=0$ and $a>0$, the azimuthal anchor-winding gradient $\partial_{\theta}\gamma$ becomes maximal at $\theta=0,\pi$ and minimal at $\theta=\pi/2,3\pi/2$. It is therefore energetically favorable for topological defects to be closer to the boundary at $\theta=0,\pi$ than at $\theta=\pi/2,3\pi/2$. Consequently, the axis of a defect pair aligns with the $x$-axis in this case (Fig.~\ref{fig2x}a, \hbox{$t=0$}). An analogous reasoning for arbitrary~$\theta_{\textrm{a}}$ implies that the boundary condition~\eqref{eq:anchwind} orients defect pairs along the axis  $\left(\cos\theta_{\textrm{a}},\sin\theta_{\textrm{a}}\right)$ when $a>0$. In the next part, we will use a dynamic generalization of the anchoring profile Eq.~(\ref{eq:anchwind}) to realize defect braiding protocols in both continuum and particle simulations.

%%%%%%%%%%%%%%%%%%
%%%%%%%%%%%%%%%%%%
\begin{figure*}%[t]
\centering
	\includegraphics[width = 1.95\columnwidth]{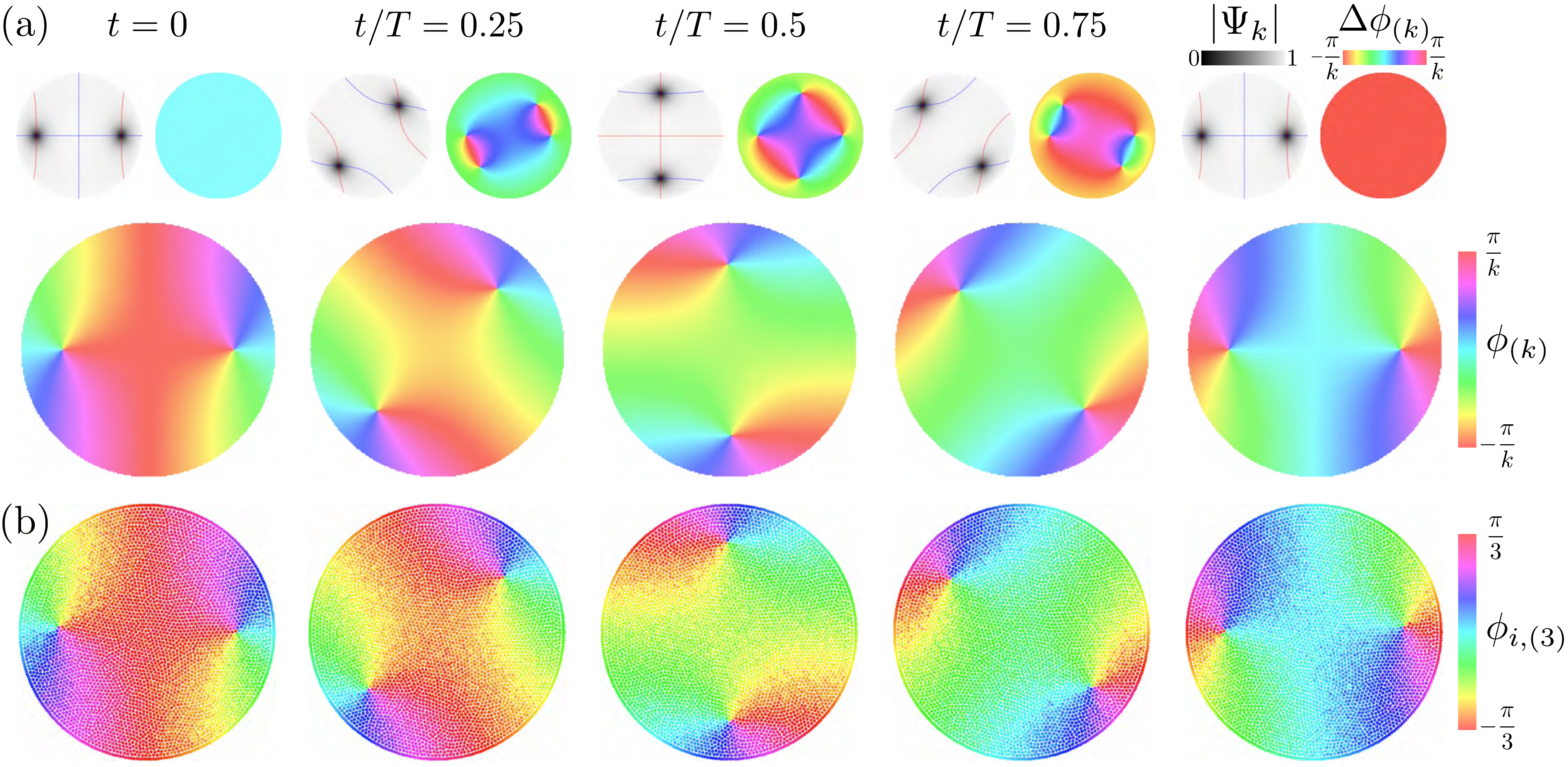}
	\caption{Demonstration of the braiding protocol in GL mean-field theory and particle simulations with aligning interactions ($g>0$) on the unit disk. (a)~A pair of identical $1/k$-defects is braided in GL simulations over the interval $t/T=[0,1]$ through a modulation of the boundary anchoring. Red and blue lines indicate $\text{Re}\Psi_k=0$ and $\text{Im}\Psi_k=0$, respectively. The $k$-atic phase shift $\Delta \phi_{(k)}$ given in Eq.~(\ref{eq:phasediff}) assumes the constant global value $\Delta \phi_{(k)}=\pi/k$ for $t=T$ (top right; Movie~4). The shown $k$-atic phases are exact stationary solutions of the GL, corresponding to perfectly adiabatic braiding with $T\rightarrow\infty$. (b)~Particle simulations ($D_rT=20$) for a braided pair of 1/3-defects replicate the GL prediction (Movie~5). Simulations used the boundary anchoring Eq.~(\ref{eq:bc1}) (mean-field) and Eq.~(\ref{eq:bcpm}) (particle model) with anchoring profile $\gamma_k$ defined through Eqs.~(\ref{eq:anchwind})--(\ref{eq:phibraiding}) with $a=0.1$, all other parameters are as in~Fig.~\ref{fig2}d.
	}
	\label{fig3}
\end{figure*}
%%%%%%%%%%%%%%%%%%
%%%%%%%%%%%%%%%%%%

\subsection{Braiding through boundary modulation}
\label{sec:stranch}

We now demonstrate how one can implement a braiding protocol that mimics the properties of anyonic states~\cite{chet08,lee19} by adiabatically changing the anchoring Eqs.~(\ref{eq:bc1}) and (\ref{eq:bcpm}) with anchoring profile $\gamma_k$ given in Eq.~\eqref{eq:anchwind} in both the mean-field description and the particle model. Specifically, we interpret defects carrying the same topological charge per Eq.~(\ref{eq:defch}) as \lq identical particles\rq\ and, accordingly, aim to braid an identical defect pair such that the phase of the final state $\Psi_k(t=T)$ differs from that of the initial state $\Psi_k(t=0)$ by a constant global shift. To this end, it is instructive to first discuss how defect braiding in the microscopic model generates an emergent anyonic exchange symmetry on a subdomain of the disk. In second step, we show how this can be extended to a protocol where defect exchanges are accompanied by global $k$-atic phase changes.

\subsubsection{Braiding-induced anyonic exchange symmetries \\in the particle model}
We consider a pair of $1/k$-defects that is initially aligned with the $x$-axis (Fig.~\ref{fig2x}a,~$t=0$). As discussed above, this configuration can be achieved by imposing the anchoring condition~\eqref{eq:bc1} with $m=2$ and $\theta_{\textrm{a}}=0$ in Eq.~\eqref{eq:anchwind}. To parametrize the exchange of defect positions, we consider the anchoring profile $\gamma_k$ given in Eq.~(\ref{eq:anchwind}) with $\theta_0=0$ and
\begin{equation}\label{eq:t0param}
\theta_{\textrm{a}}(t)=\frac{\pi t}{T}.
\end{equation} 
For increasing $t\in[0,T]$ with $T>0$, this leads to a counter-clockwise rotation of the preferred localization axis of the defect pair during which the two defects move along complementary semi-circles (see $|\psi_3|$ in Fig.~\ref{fig2x}a). These trajectories eventually enclose a sub-domain where particle orientation differences fluctuate around \hbox{$\Delta\alpha_i=\alpha_i(T)-\alpha_i(0)=2\pi/k$} (shown in Fig.~\ref{fig2x}a for \hbox{$k=3$}), corresponding to an anyonic exchange symmetry~\cite{lein77,wilc82,kita03,chet08,lee19} of the \textit{polar} order parameter field~$\psi_1$~[Eq.~(\ref{eq:psikdef})]. The emergence of this sub-domain can be understood as follows: Short-range alignment interactions together with the boundary anchoring $\alpha_i|_{\partial S}$ essentially fix orientations of particles near the boundary. Consequently, a passing-by branch cut led by a $1/k$-defect causes a change of the interior local particle orientations by $2\pi/k$, as shown for $k=3$ in Fig.~\ref{fig2x}b. 

Braiding clockwise instead, $t\in[0,-T]$, changes local particle angles by $-2\pi/k$ when defects pass by and consequently leads to a sign flip of~$\Delta\alpha_i$ (Fig.~\ref{fig2x}c), analog to the properties of an anyonic exchange symmetry~\cite{chet08}. Finally, by performing $k$ such braids, we demonstrate that the orientation angle change $\Delta\alpha_i$ after each braid is quantized in steps of $\pm2\pi/k$, despite the stochasticity in the particle model (Fig.~\ref{fig2x}c). Indeed, three consecutive braids of $3$-atic particles return all particles in the subdomain enclosed by the defect trajectories to their initial orientations~(Movie~3).

\subsubsection{Defect braiding and global k-atic phase changes in mean-field theory and particle model}
Our next goal is to identify a boundary anchoring protocol for which defect braiding is accompanied by a \textit{global} $k$-atic phase shift. To this end, also particle orientations in the outer annulus, where so far \hbox{$\Delta\alpha_i=0$}~(Fig.~\ref{fig2x}a,c), have to change with the braiding. Adding a constant global angle $\mp2\pi/k$ to the boundary anchoring profile is not sufficient, as it equally affects the whole disk such that the angle difference $\Delta\alpha_i=\pm2\pi/k$ between annulus and the domain enclosed by the defect trajectory remains. However, a constant global phase shift of $k$-atic director orientations $\phi_{i,(k)}$ [Eq.~(\ref{eq:micrangmap})] and of the $k$-atic mean-field phase $\phi_{(k)}$ [Eq.~(\ref{eq:phasedef})] can be achieved by exploiting the fact that for both quantities $-\pi/k$ is identified with~$\pi/k$. Specifically, we use the anchoring profile Eq.~(\ref{eq:anchwind}) with $\theta_{\textrm{a}}(t)$ given in Eq.~(\ref{eq:t0param}) and
\begin{align}
\label{eq:phibraiding}
\theta_0(t)=\frac{\pi t}{2T}.
\end{align}
As before, this protocol induces a counter-clockwise braid that exchanges the two identical defects. However, now Eq.~(\ref{eq:phibraiding}) simultaneously modulates the $k$-atic phase~$\phi_{(k)}$ within the annulus and the sub-domain enclosed by defect trajectory by~$-\pi/k$ such that a globally constant $k$-atic phase difference 
\begin{equation}\label{eq:phasediff}
\Delta\phi_{(k)}=\frac{1}{k}\text{arg}\left[\Psi_k(T)\Psi^*_k(0)\right]=\frac{\pi}{k}
\end{equation}
emerges, as illustrated by stationary solutions of the GL equation for different $t/T$ in Fig.~\ref{fig3}a (Movie~4). The stochastic particle model undergoes a noisy realization of this texture sequence (Fig.~\ref{fig3}b, Movie~5) if the same boundary anchoring~$\gamma_k$ is used in Eq.~(\ref{eq:bcpm}). 

More generally, this demonstrates how one can implement an exchange of two identical defects with an accompanying global $k$-atic phase change in the complex order-parameter field of DLCs. We emphasize, however, that a global phase change in the k-atic director field necessitates a corresponding phase change at the boundary.

%%%%%%%%%%%%%%%%%%%%%%%%%%%%
%%%%%%%%%%%%%%%%%%%%%%%%%%%%

\section{Spontaneous chiral symmetry breaking and braiding in anti-aligning DLC\lowercase{s} ($g<0$)}
\label{sec:SH}
%%%%%%%%%%%%%%%%%%%%%%%%%%%%
%%%%%%%%%%%%%%%%%%%%%%%%%%%%

Having focused on aligning interactions in the previous section, we now consider $k$-atic DLCs with short-range anti-aligning interactions ($g<0$; Fig.~\ref{fig1}a,b). In this case, the mean-field model in Eq.~\eqref{eq:RGLgen} takes the form of a Swift-Hohenberg (SH) equation for the  complex order parameter~$\Psi_k$.  For boundary anchorings that impose a topological charge of $1/k$ on a unit disk, the SH equation predicts a spontaneous chiral symmetry breaking of texture patterns that is also observed in the microscopic particle model. We then show that his chiral symmetry breaking  can be understood analytically by constructing a stationary solution to the linearized complex SH equation. Last but not least, to demonstrate the versatility of the braiding protocol from Sec.~\ref{sec:stranch}, we will perform an braiding operation for a defect pair in anti-aligning DLCs in both SH equation and microscopic model.

%%%%%%%%%%%%%%%%%%%%%%%%%%%%%%
\subsection{SH mean-field theory for anti-aligning $k$-atics}
%%%%%%%%%%%%%%%%%%%%%%%%%%%%%%
We consider the microscopic model Eq.~(\ref{eq:micrmod}) with \mbox{$g<0$}, which favors anti-aligning configurations of nearby $k$-atic particle directors. For a small but finite interaction range $R_\alpha>0$, the mean-field dynamics of the $k$-atic mode can be approximated by~(Appendix~\ref{app:CG}) 
\begin{equation}\label{eq:LinDynGen}
\bar{\tau}\partial_t\psi_k\approx h_k-\left(\beta_1R_{\alpha}^2\nabla^2+\beta_2R_{\alpha}^4\nabla^2\nabla^2\right)\psi_k,
\end{equation}
where the homogeneous terms $h_k$ were defined previously in Eq.~(\ref{eq:kcoarse}). The coefficients $\beta_1$ and $\beta_2$ in Eq.~(\ref{eq:LinDynGen}) depend on the spatial interaction kernel of the microscopic model. Assuming, as before, equally weighted interactions between particles within a neighborhood of radius~$R_{\alpha}$, the kernel determines the dispersion relation for perturbations of the $k$-atic mode $\psi_k\sim f_k$ around the disordered state $f_k=0$, which is approximated by the parameters $\beta_1$~and~$\beta_2$~(Appendix~\ref{appsec:psdiffop}).

The coarse-graining result Eq.~(\ref{eq:LinDynGen}) implies a mean-field model for anti-aligning DLCs  of the general form Eq.~(\ref{eq:RGLgen}) with 
\begin{equation}\label{eq:opSHmodel}
\mathcal{L}(\nabla^2)=-L_1^2\nabla^2-L_2^4\nabla^2\nabla^2,
\end{equation}
corresponding to a SH equation~\cite{cross09} for the complex order parameter $\Psi_k$. Equation~(\ref{eq:LinDynGen}) specifies the mean-field parameters $L_1$ and $L_2$ in Eq.~(\ref{eq:opSHmodel}) in terms of the interaction radius $R_{\alpha}$ as
\begin{equation}\label{eq:L1L2def}
L_1\simeq\beta_1^{1/2}R_{\alpha}\hspace{1cm}L_2\simeq\beta_2^{1/4}R_{\alpha}.
\end{equation}
The relaxation dynamics corresponding to such a mean-field theory is generated by a Landau-Brazovskii energy~$\mathcal{E}_k$ for the complex order parameter~$\Psi_k$~\cite{braz75}~(Appendix~\ref{app:freeEnSH}). Furthermore, the ratio $L_2^2/L_1$ signals an emergent mesoscopic length-scale in the phase field, arising from the competition between anti-aligning particle interactions. In particular, we expect for the dynamic Eq.~(\ref{eq:RGLgen}) with $\mathcal{L}(\nabla^2)$ given in Eq.~(\ref{eq:opSHmodel}) finite wavelength instabilities at wavevector amplitude $q_0^2$, when $A<A_*$, where
\begin{equation}
q_0^2=\frac{L_1^2}{2L_2^4}\hspace{1cm}A_*=\frac{L_1^4}{4L_2^4},\label{eq:q0A}
\end{equation}
a prediction that is validated below in both the particle model in the quantitatively mapped complex SH equation.

%%%%%%%%%%%%%%%%%%%%%%%%%%%%%%
\subsection{Spontaneous chiral symmetry breaking of texture patterns}\label{sec:symbr}
%%%%%%%%%%%%%%%%%%%%%%%%%%%%%%

Even when microscopic particles are achiral, the interplay between anti-aligning particle interactions and boundary conditions can give rise to an interesting spontaneous chiral symmetry breaking phenomenon. In the following, this is first demonstrated using the nonlinear dynamics of 3-fold symmetric particles described by $k=3$ and $g<0$ in Eq.~(\ref{eq:micrmod}) and comparing them with predictions from the corresponding SH mean-field Eq.~\eqref{eq:RGLgen} with operator Eq.~(\ref{eq:opSHmodel}). Finally, we describe analytic stationary solutions of the linearized complex SH equation that recapitulate the observed patterns, as well as the bifurcation into the chiral symmetry breaking transition.

%%%%%%%%%%%%%%%%%%
%%%%%%%%%%%%%%%%%%
\begin{figure*}%[t]
\centering
	\includegraphics[width = 1.95\columnwidth]{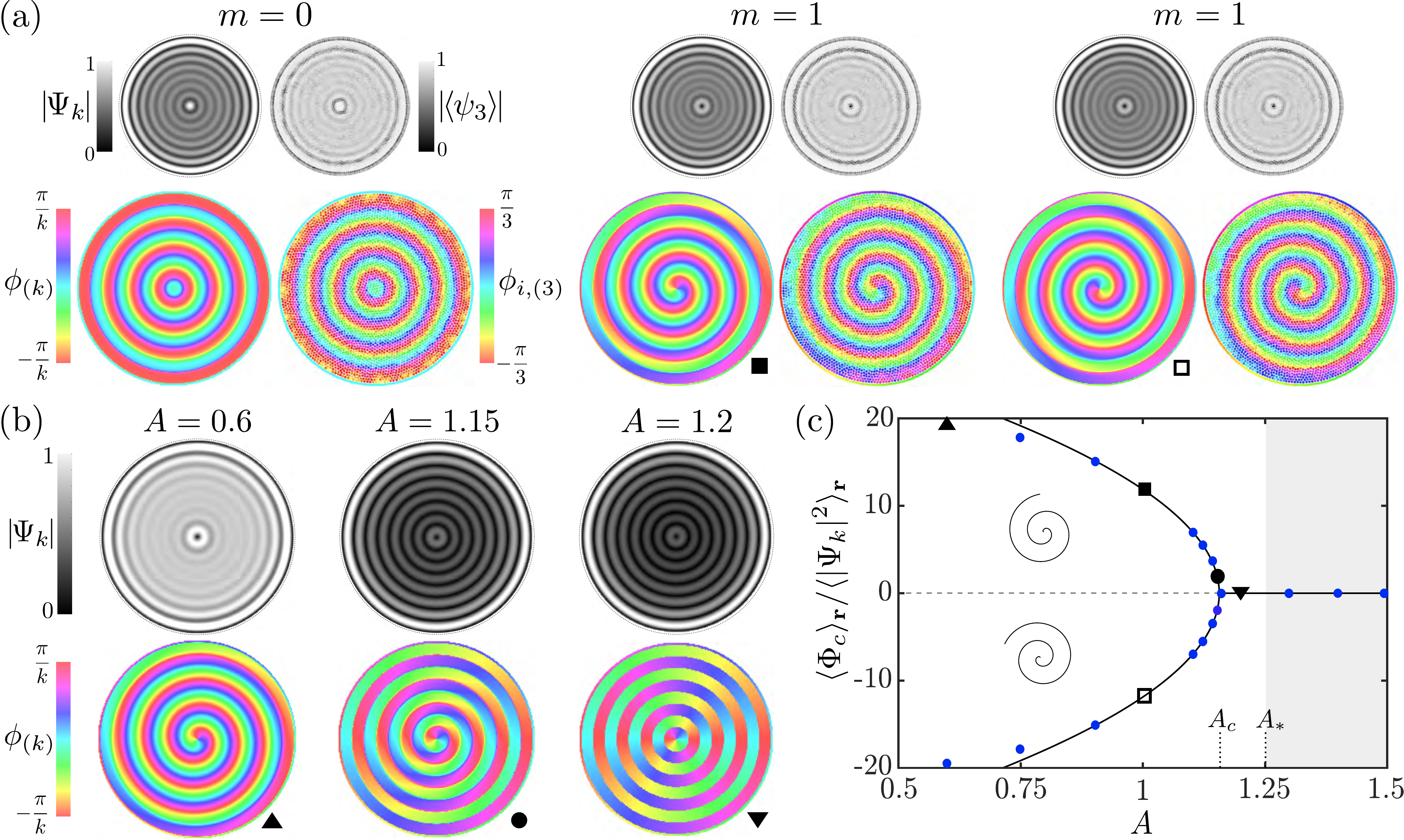}
	\caption{Chiral symmetry breaking in phase field textures of the SH mean-field theory and particle simulations with anti-aligning interactions ($g<0$) on a unit disk. (a)~Stationary order-parameters (top) and phase-field solutions (bottom) for the SH mean-field Eq.~(\ref{eq:RGLgen}) with $\mathcal{L}=-L_1^2 \nabla^2 - L_2^4 (\nabla^2)^2$ and particle model Eq.~(\ref{eq:micrmod}) with anti-aligning interactions using the boundary anchorings Eq.~(\ref{eq:bc1}) and Eq.~(\ref{eq:bcpm}), respectively, with $\gamma_k=\theta$. A stationary defect-free axisymmetric state exists for topologically trivial boundary anchoring with $m=0$. For $m=1$, chiral textures with $1/k$-defect at the center form spontaneously in both mean-field and particle simulations (Movie~6).  Mean-field simulation parameters were  $A=B=1$,  $L_1$ and $L_2$ as defined in Eq.~(\ref{eq:L1L2def}) with $\beta_1=0.1$, $\beta_2=0.002$ (characteristic wavenumber $q_0=25$) and $R_{\alpha}=0.2$.  Particle simulation parameters:  $k=3$, $g=-1$, and all other parameters as in Fig.~\ref{fig2}d. 
	(b)~Examples of spontaneously formed stationary solution of the complex SH equation for $m=1$ and selected values of the control parameter $A$ with parameters identical to those in panel (a). 
	(c)~The phase-chirality parameter~$\Phi_c$ defined in Eq.~(\ref{eq:chphasepara}) characterizes the spontaneous symmetry-breaking into chiral textures of different handedness ($\langle\cdot\rangle_{\mathbf{r}}$ denotes a spatial averages). The bulk dynamics of the complex SH equation is linearly stable for $A>A_*$ (gray-shaded region), but achiral ring patterns remain due to the boundary anchoring Eq.~(\ref{eq:bc1}) with $|\Psi_k|_{\partial S}=1$.  Phase-chirality parameters measured in additional simulations for different values of $A$  (small blue dots) approach the curve $\sim\pm\sqrt{A_c-A}$ for $A\searrow A_c$ with $A_c\approx1.155$, consistent with a supercritical pitchfork bifurcation.}
	\label{fig4}
\end{figure*}
%%%%%%%%%%%%%%%%%%
%%%%%%%%%%%%%%%%%%

\subsubsection{Pattern formation on the unit disk}
We first consider  a topologically trivial boundary anchoring corresponding to  $m=0$ in Eq.~(\ref{eq:bc1}). In this case,  the SH mean-field theory predicts the existence of defect-free azimuthally symmetric stationary states that can indeed be observed in simulations of the anti-aligning particle model~(Fig.~\ref{fig4}a, $m=0$). However, although such defect-free states are long-lived in the presence of noise, they only form from suitably pre-patterned initial conditions~(Appendix~\ref{app:NumMeth}), whereas random initial conditions typically lead to ring-shaped patterns that are intersected by chains of $\pm 1/k$-point defects with zero topological net charge~(Fig.~\ref{figS2}c). 
\par

For a boundary anchoring Eq.~(\ref{eq:bc1}) with $\gamma_k=\theta$ and $m=1$, the SH mean-field theory predicts the spontaneous formation of a spiral-shaped chiral texture with a $1/k$-defect at the center of the disk, while the $k$-atic order parameter $|\Psi_k|$ maintains an azimuthal symmetry. Again, the particle model confirms this prediction~(Fig.~\ref{fig4}a, right and Movie~5). Such chiral textures are reminiscent of spiral patterns that have recently been observed theoretically and experimentally in cholesteric LCs~\cite{poll19,tran20}. However, the latter arise from chiral microscopic interactions, while spiral texture patterns in our system arise from a spontaneous symmetry breaking among isotropically interacting achiral particles.

%%%%%%%%%%%%%%%%%%
%%%%%%%%%%%%%%%%%%
\begin{figure*}%[t]
\centering
	\includegraphics[width = 2\columnwidth]{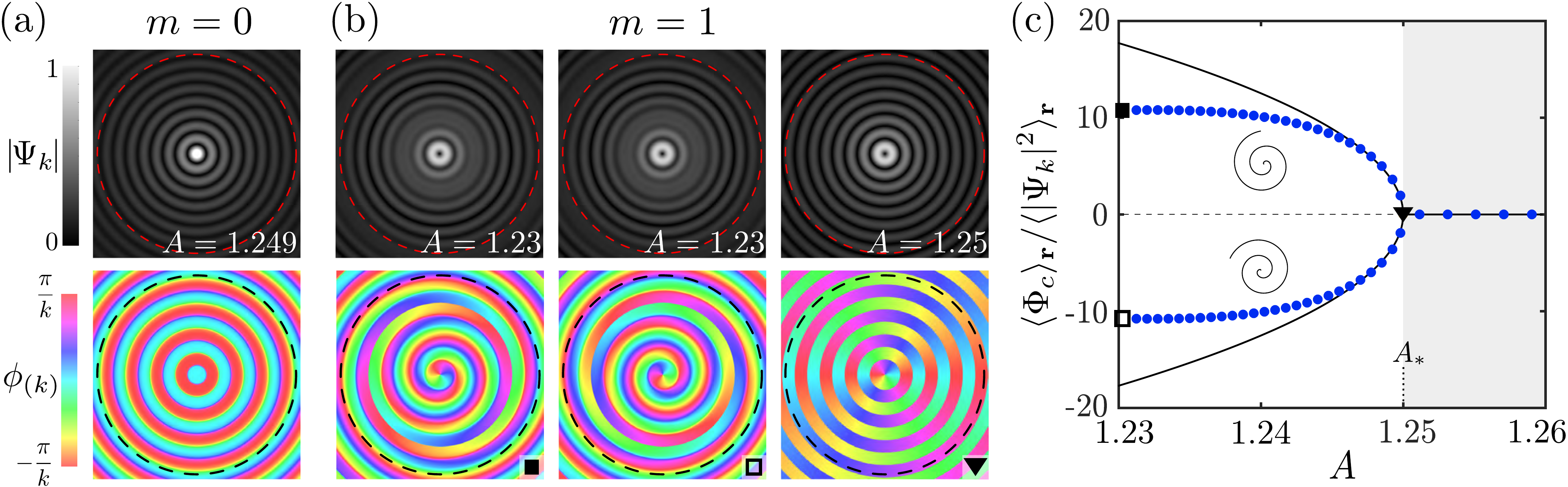}
	\caption{Analytic solutions of the linearized complex SH equation on infinite domains explain pattern formation and chiral symmetry breaking. (a)~Defect-free solution (Eq.~(\ref{eq:complSHsolmain}) with $m=0$ and \smash{$\mu_0=\nu_0^*=e^{i\pi/4}$}) recapitulates the wavelength-doubling between amplitude and phase patterns~(Fig.~\ref{fig4}a, Appendix~\ref{app:wldiff}). (b)~Examples from the family of analytic solutions Eq.~(\ref{eq:complSHsolmain}) with $m=1$, \smash{$\mu_1=e^{i\pi/4}$} and \smash{$\nu_1=\pm e^{-i\pi/4}$} (\lq$+$\rq:~right-handed spirals, \lq$-$\rq:~left-handed spirals) show bifurcation into chiral texture patterns, as quantified in (c), when the control parameter~$A$ becomes smaller than the critical value $A_*$ [see Eq.~(\ref{eq:q0A})]. All fields in (a) and (b) use the same length scales $L_1$ and $L_2$ as in Fig.~\ref{fig4} and are compatible with the director anchoring Eq.~(\ref{eq:bc1}) for $\gamma_k=\theta$ at a unit disk boundary (red and black dashed lines). (c)~Phase-chirality with $\Phi_c$ defined in Eq.~(\ref{eq:chphasepara}), evaluated on a unit disk centered at $r=0$ for analytic solutions (blue dots). Black symbols indicate exemplary solutions shown in~(b). Solid black lines depicts a fit $\sim\sqrt{A_*-A}$ near the critical value $A_*$. Compared to these analytic solutions, numerical solutions of the fully nonlinear SH equation on a finite domain exhibit a small shift in the critical value $A_c$ of the bifurcation (Fig.~\ref{fig4}c, $A_c<A_*$), reflecting additional effects from the nonlinearity $\sim B|\Psi_k|^2\Psi_k$ and from boundary conditions.}
\label{figtheo}
\end{figure*}
%%%%%%%%%%%%%%%%%%
%%%%%%%%%%%%%%%%%%

\subsubsection{Characterization of the chiral symmetry breaking~transition}
Upon varying~$A$ in simulations of the complex SH equation, while keeping all other parameters fixed, we find that textures become chiral only below a critical value~$A_c$, but remain azimuthally symmetric if $A>A_c$ (Fig.~\ref{fig4}b,c). To quantify this transition, we introduce the phase-chirality parameter~(Appendix~\ref{app:chirsol})
\begin{equation}\label{eq:chphasepara}
\Phi_c=k|\Psi_k|^2\mathbf{e}_r\cdot\nabla\phi_{(k)}.
\end{equation}
This quantity characterizes chiral signatures in textures around a defect by measuring the radial contributions in the k-atic phase gradient. In particular, $\Phi_c>0$ ($\Phi_c<0$) indicates in our system the presence of right-handed (left-handed) texture spirals leading into the defect~(Fig.~\ref{fig4}c). Averaging $\Phi_c$ across the disc domain, we observe a continuous transition from achiral to chiral textures at $A_c$. The transition exhibits the characteristics of a supercritical pitchfork bifurcation with left- and right-handed texture patterns forming with equal probability from random the initial conditions~(Fig.~\ref{fig4}c). Note, that the transition point $A_c$ tends on an infinite domain to the linear stability threshold $A_*$ (see~Sec.~\ref{sec:SHanalyt}).

\subsubsection{Wavelength-doubling in amplitude and phase patterns}\label{sec:wldoub}

For all stationary solutions shown in Fig.~\ref{fig4}a, the wavelength of texture patterns is in quantitative agreement with the mean-field prediction~\mbox{$\lambda_0=2\pi/q_0\approx0.25$}. This can be seen from the $k$-atic phases $\phi_{(k)}$ and $\phi_{i,(3)}$ that cycle about $1/\lambda_0\approx 4$ times through $2\pi/k$ between the boundary and center of the unit disk. Interestingly, the wavelength of texture patterns is twice as large as the wavelength of order parameter amplitude patterns in both the particle model ($|\langle\psi_3\rangle|$) and the mean-field theory ($|\Psi_k|$). In the particle model, this can be rationalized as follows: Two single anti-aligning particles interacting via Eq.~(\ref{eq:micrmod}) ($g<0$) are stationary for a director angle difference of $\pi/k$. However, when anti-aligning interactions are present over a finite range and include several particles, it becomes energetically favorable to form finite-sized regions, each with an approximately constant director orientation (corresponding to high orientational order $|\langle\psi_3\rangle|$), but with a difference of $\pi/k$ to director orientations in the directly neighboring region. As a result, one cycle of the $k$-atic phase through $2\pi/k$ contains two regions of high order and consequently the wavelength of amplitude patterns is half the wavelength of texture patterns. On the unit disk, regions of high order form annuli due to the boundary anchoring (Figs.~\ref{fig4}~and~\ref{figS2}c), whereas on periodic domains they are given by checkerboard patterns~(Fig.~\ref{figS3}). The mean-field theory recapitulates these non-trivial consequences of microscopic anti-alignment in all cases with good quantitative agreement, which can be understood from analytic stationary solutions of the linearized complex SH equation.

%%%%%%%%%%%%%%%%%%
%%%%%%%%%%%%%%%%%%
\begin{figure*}%[t]
\centering
	\includegraphics[width = 1.95\columnwidth]{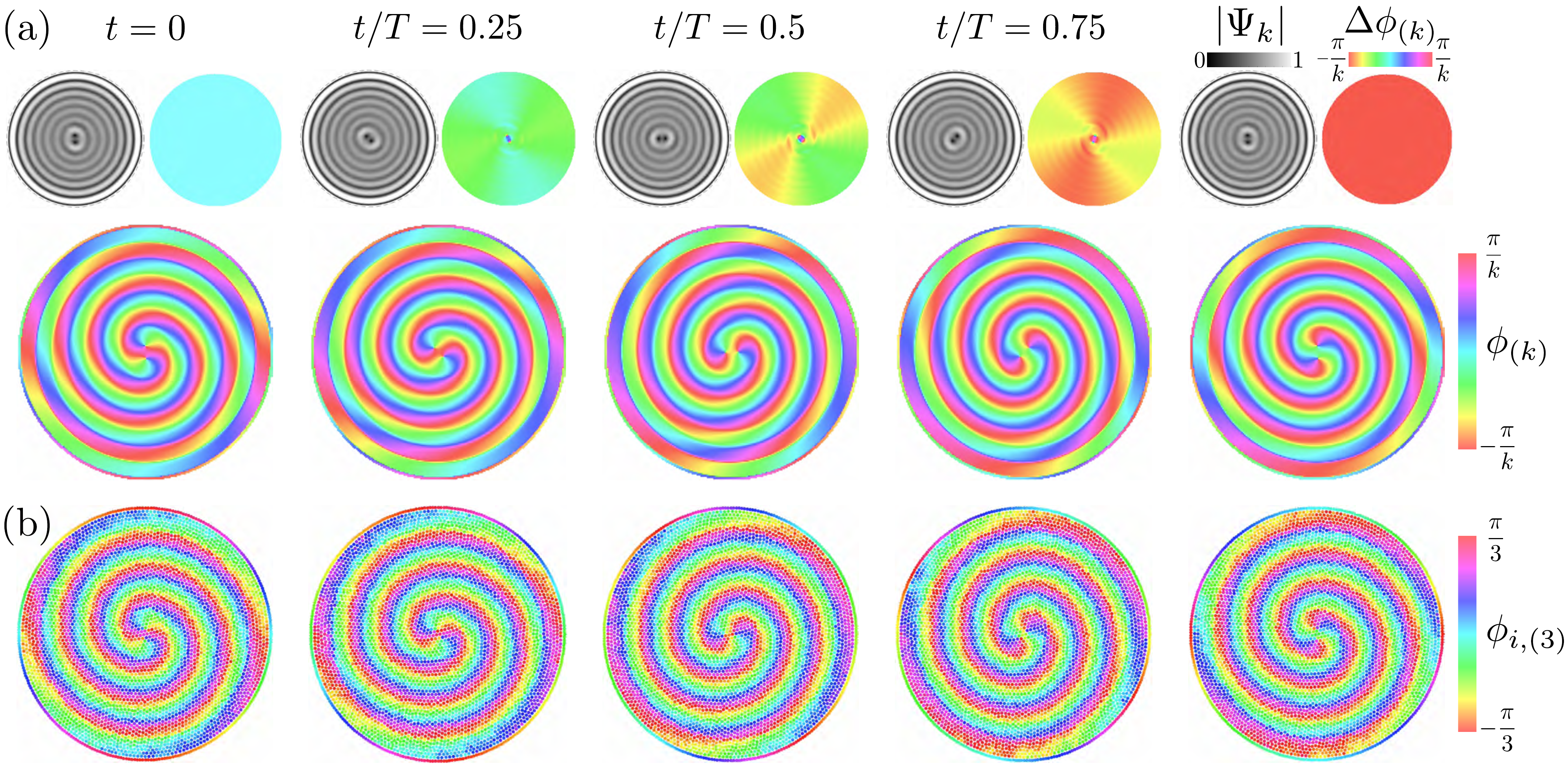}
	\caption{Demonstration of the braiding protocol in (a) the SH mean-field theory and (b) particle simulations with anti-alignment interactions ($g<0$) on the unit disk (Movies~7 and 8). Boundary anchorings Eq.~(\ref{eq:bc1}) (mean-field) and Eq.~(\ref{eq:bcpm}) (particle model) with $m=2$ impose a total defect charge of $2/k$ on the domain, supporting two stable $1/k$ point defects near the disc center and outwards spiraling textures. Similar to Fig.~\ref{fig3}, we applied the boundary-controlled braiding protocol with $\gamma_k$ defined through Eqs.~\eqref{eq:anchwind}--(\ref{eq:phibraiding}) and $a=0.2$, while using the model parameters from Fig.~\ref{fig4}. 
	The $k$-atic mean-field phase fields in (a) are exact stationary solutions of the complex SH equation, corresponding to perfectly adiabatic braiding with $T\rightarrow\infty$.  
	In both mean-field and particle simulations, the initial defect-pair state does not form spontaneously from random initial conditions, but can be robustly generated as a stationary state from initial conditions that are sufficiently close to a double-spiral texture~(Appendix~\ref{app:NumMeth}). The braiding protocol is started at $t=0$ by modulating the boundary anchoring as described in Sec.~\ref{sec:stranch}.  At the end of the braiding process, the two point defects have exchanged their position and the director field has acquired a constant global $k$-atic phase shift~$\Delta\phi_{(k)}=\pi/k$ in the SH equation (a) and of $\pi/3$ in the anti-aligning particle model with $k=3$ (b).}
\label{fig5}
\end{figure*}
%%%%%%%%%%%%%%%%%%
%%%%%%%%%%%%%%%%%%

\subsubsection{Analytic solutions of the complex SH equation}\label{sec:SHanalyt}

All observations from Sec.~\ref{sec:symbr} can be recapitulated by analytic stationary solutions near the linear instability. Specifically, we show in Appendix~\ref{app:ansolSH} that exact solutions of the linearized complex SH equation 
$$
A\Psi_k+L_1^2\nabla^2\Psi_k+L_2^4\nabla^2\nabla^2\Psi_k=0
$$ 
can be found by solving the equivalent bi-Helmholtz equation
\begin{equation}\label{eq:biHelmmain}
\left(\nabla^2+q_+^2\right)\left(\nabla^2+q_-^2\right)\Psi_k=0,
\end{equation}
where~\mbox{$q_{\pm}^2=q_0^2\pm\sqrt{\Delta A}/L_2^2$}~with~\mbox{$\Delta A=A_*-A$} (Appendix~\ref{app:ansolSH}), and $q_0$ and $A_*$ are given in Eq.~(\ref{eq:q0A}). Solutions of Eq.~(\ref{eq:biHelmmain}) in polar coordinates $(r,\theta)$ take the form
\begin{equation}\label{eq:complSHsolmain}
\Psi_k(r,\theta)=\sum_{m=0}^{\infty}\left[\mu_mJ_m(q_-r)+\nu_mJ_m(q_+r)\right]e^{im\theta},
\end{equation}
where $J_m(x)$ are Bessel functions of the first kind, and $\mu_m$ and $\nu_m$ denote complex integration constants. Each mode $m$ in Eq.~(\ref{eq:complSHsolmain}) represents an azimuthally symmetric amplitude pattern $|\Psi_k|$ that harbors a defect of topological charge $q_\text{d}=m/k$ at $r=0$, consistent with stationary states shown in Fig.~\ref{fig4}a,b.

At the critical point $A=A_*$, where $q_{\pm}=q_0$, all modes in Eq.~(\ref{eq:complSHsolmain}) are of the form $\Psi_k\sim J_{m}(q_0r)e^{im\theta}$. Such fields represent concentric annuli of phase patterns with wavelength $\sim2\pi/q_0$ and amplitude patterns with half this wavelength, $\sim\pi/q_0$~(Fig.~\ref{figtheo}a,b, Appendix~\ref{app:wldiff}). Beyond the critical point $A<A_*$, where $q_+\ne q_-$, the mode $m=1$ in Eq.~(\ref{eq:complSHsolmain}) can represent right- and left-handed chiral texture patterns $\phi_{(k)}$~(Fig.~\ref{figtheo}b, Appendix~\ref{app:chirsol}), while amplitude patterns~$|\Psi_k|$ maintain azimuthal symmetry. These key features in both achiral and chiral stationary analytic solutions recapitulate our findings from numerical solutions of the fully nonlinear mean-field theory~(Fig.~\ref{fig4}), including the pitchfork bifurcation of stationary solutions beyond a critical control parameter threshold $A_*$ into chiral texture patterns~(Fig.~\ref{figtheo}c).

%%%%%%%%%%%%%%%%%%%%%%
\subsection{Braiding of chiral texture patterns}
%%%%%%%%%%%%%%%%%%%%%%%

Similar to their achiral counterparts, the $k$-atic fields describing chiral defect pairs in anti-aligning DLCs can also be the braided by applying the adiabatic protocol from Sec.~\ref{sec:stranch}. To demonstrate this, we use in both SH mean-field and particle simulations the boundary condition Eq.~(\ref{eq:bc1}) with $m=2$ and fix the anchoring profile~$\gamma_k$ from  Eq.~(\ref{eq:anchwind}) with $a=0.2$. This boundary condition stabilizes now a vertically oriented pair of $1/k$ defects near the disk center, which is surrounded by texture spirals that intertwine towards the boundary (Fig.~\ref{fig5}, $t=0$). To prepare this initial state in the continuum and particle simulations, one can let the system relax from  suitably pre-patterned  textures~(Appendix~\ref{app:NumMeth}). The resulting two-defect state is stable and can be robustly braided using the anchoring profile parametrization given in Eq.~(\ref{eq:phibraiding}).  Snapshots of the braiding sequences, selected from Movies~7 and 8, are shown in~Fig.~\ref{fig5}. At the end of the braiding operation, the defects have exchanged their positions and the complex order parameter field $\Psi_k$ has acquired a constant global phase shift $\Delta\phi_{(k)}=\pi/k$~[Eq.~(\ref{eq:phasediff})].

\section{Conclusions}

The above analysis shows that the relaxation behavior and adiabatic manipulation of 2D liquid crystals composed of $k$-fold symmetric particles can be accurately described within a unified mean-field theory for a complex order-parameter field. Due to the generic character of the underlying particle model, which merely assumed overdamped short-range $XY$-type interactions on a disordered lattice, we expect that the above ideas can be experimentally implemented and tested in different ways. Promising candidates include colloidal systems~\cite{liu16,nied17} with predefined symmetries and controllable steric~\cite{zhao12,vutu14,wang14,li16,loff18}, magnetic~\cite{grzy00,soni19} or chemical~\cite{yi13,andr13} interactions. The main experimental challenge will be to enforce the required short-range orientational interactions while simultaneously suppressing positional order. This could, for example, be achieved using weakly multidisperse $k$-atic platelet systems, similar to those realized in Ref.~\cite{zhao12,loff18}. Other promising candidate systems could be thin films of 3-fold symmetric molecules~\cite{bowi17} or $k$-fold symmetric DNA-origami structures~\cite{Chao:2018aa,Veneziano1534}, building on recently developed experimental techniques~\cite{Siavashpouri:2017aa} for the assembly and control of DNA-origami-based liquid crystals.
\par
From a general theoretical perspective, $k$-atic DLC systems~\cite{bowi09} provide a useful classical framework for studying and visualizing fractional topological excitations and their exchange properties.  Since the energetic correspondence with  quantum fluids only holds at the mean-field level, it remains an interesting open question whether and how the statistical properties of fractional defects in DLCs depend on their braiding behavior. In addition, the above results suggest multiple directions for future research, including generalizations to passive and active $k$-atic hydrodynamic systems~\cite{wens13,giomi21} in two and three dimensions, which can be expected to exhibit new forms of energy transport and turbulence~\cite{giomi15,alert20}.

\section*{Acknowledgements}
We thank Vili Heinonen, Martin Zwierlein and Mehran Kardar for helpful discussions and insightful comments. This work was supported by a Longterm Fellowship from the European Molecular Biology Organization (EMBO ALTF~528-2019, A.M.), a Postdoctoral Research Fellowship from the Deutsche Forschungsgemeinschaft (DFG Project~431144836, A.M.), a Complex Systems Scholar Award from the James S. Mc-Donnell Foundation (J.D.) and the Robert E. Collins Distinguished Scholarship Fund (J.D.).

\bibliography{References}

\clearpage

\appendix

\renewcommand\thefigure{S\arabic{figure}}    
\setcounter{figure}{0} 

\begin{center}
\textbf{APPENDIX}
\end{center}

%%%%%%%%%%%%%%%%%%
%%%%%%%%%%%%%%%%%%
\begin{figure*}%[t]
\centering
	\includegraphics[width = 1.95\columnwidth]{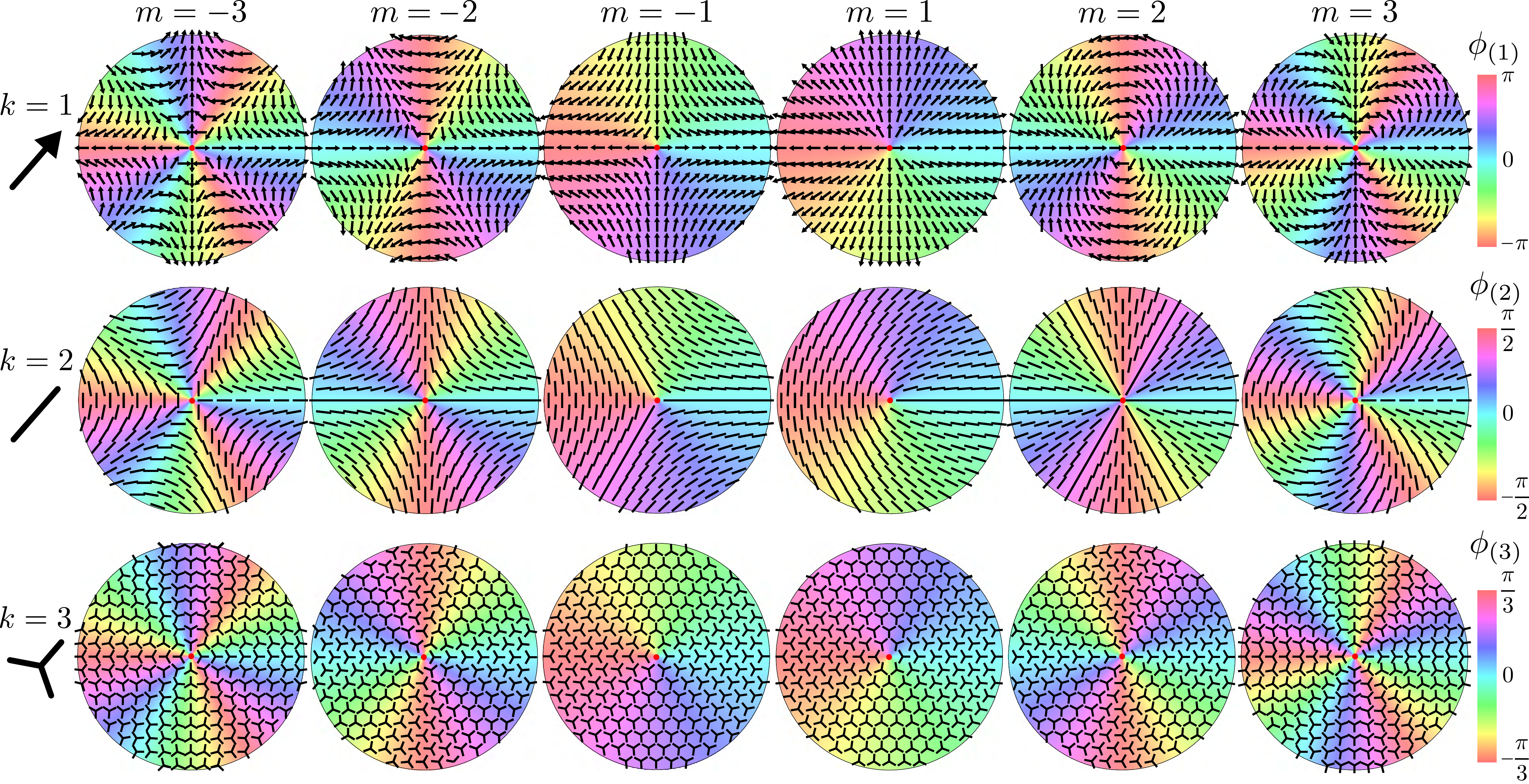}
	\caption{Visualization of defects formed by polar ($k=1$), nematic ($k=2$) and tri-atic $(k=3)$ particles and their $k$-atic phases~$\phi_{(k)}$ for varying winding numbers $m$. The topological charge defined in Eq.~(\ref{eq:defch}) of each of the defects is $q_\text{d}=m/k$.}
\label{figS0}
\end{figure*}
%%%%%%%%%%%%%%%%%%
%%%%%%%%%%%%%%%%%%

%%%%%%%%%%%%%%%%%%%%%%%%%%%%%%%%%%%%%%%%
%%%%%%%%%%%%%%%%%%%%%%%%%%%%%%%%%%%%%%%%
\section{Coarse-graining of the microscopic model}
\label{app:CG}
%%%%%%%%%%%%%%%%%%%%%%%%%%%%%%%%%%%%%%%%
%%%%%%%%%%%%%%%%%%%%%%%%%%%%%%%%%%%%%%%%
We describe the coarse-graining of the minimal microscopic model Eq.~(\ref{eq:micrmod}) and show how spatial interactions with finite-range range $R_{\alpha}$ give rise to the operators $\mathcal{L}=R_{\alpha}^2\nabla^2/8$ and $\mathcal{L}=-\beta_1R_{\alpha}^2\nabla^2-\beta_2R_{\alpha}^4(\nabla^2)^2$ in the coarse-grained dynamics Eqs.~(\ref{eq:kcoarseany}) and (\ref{eq:LinDynGen}), respectively.

%%%%%%%%%%%%%%%%%%%%%%
\subsection{Hierarchy of mode equations and linearization}
\label{app:CG-1}
%%%%%%%%%%%%%%%%%%%%%
We follow the standard coarse-graining approach~\cite{dean96,bert09,farr12}, by using It\^{o} calculus, neglecting multiplicative noise terms and factorizing pair correlations, to derive a dynamic equation for the one-particle probability density function $f(\alpha,\mathbf{r},t)$ from the microscopic model Eq.~(\ref{eq:micrmod}). For a general interaction kernel $\hat{I}(\mathbf{r})$ that describes how the orientational interactions with neighboring particles are spatially weighted, this equation takes the form
\begin{widetext}
\begin{equation}\label{eq:cg1}
\partial_tf(\mathbf{r},\alpha,t)=\frac{g}{\pi R_{\alpha}^2}\partial_{\alpha}\iint d\alpha'd\mathbf{r}'f(\mathbf{r},\alpha,t)f(\mathbf{r}',\alpha',t)\sin\left[k\left(\alpha-\alpha'\right)\right]\hat{I}(\mathbf{r}-\mathbf{r}')+D_r\partial_{\alpha}^2f.
\end{equation}
\end{widetext}
It is convenient to define an effective, normalized interaction kernel as $I(\mathbf{r})=\hat{I}(\mathbf{r})/(\pi R_{\alpha}^2)$. Using the \hbox{Fourier-representations}
\begin{align}
f(\mathbf{r},\alpha,t)&=\frac{1}{(2\pi)^3}\sum_{n\in\mathbb{Z}}\int d\mathbf{q}\,\tilde{f}_n\left(\mathbf{q},t\right)e^{-i\left(n\alpha+\mathbf{r}\cdot\mathbf{q}\right)}\\
I(\mathbf{r})&=\frac{1}{(2\pi)^2}\int d\mathbf{q}\,\tilde{I}\left(\mathbf{q}\right)e^{-i\mathbf{r}\cdot\mathbf{q}}\label{eq:Ir}
\end{align}
in Eq.~(\ref{eq:cg1}), we can find a hierarchy of coupled dynamic equations for the modes
\begin{equation}
f_n(\mathbf{r},t)= \frac{1}{(2\pi)^2}\int d\mathbf{q}\,\tilde{f}_n\left(\mathbf{q},t\right)e^{-i\mathbf{r}\cdot \mathbf{q}}.
\end{equation}
For $k$-atic interactions as given in the microscopic model Eq.~(\ref{eq:micrmod}), this hierarchy takes the form
\begin{widetext}
\begin{equation}
\partial_tf_n(\mathbf{r},t)=\frac{gn}{8\pi^2}\int d\mathbf{q}\left[f_{n-k}\left(\mathbf{r},t\right)\tilde{f}_k(\mathbf{q},t)-f_{n+k}\left(\mathbf{r},t\right)\tilde{f}_{-k}\left(\mathbf{q},t\right)\right]\tilde{I}\left(\mathbf{q}\right)e^{-i\mathbf{r}\cdot\mathbf{q}}-D_rn^2f_n\left(\mathbf{r},t\right),\label{eq:gensys}
\end{equation}
\end{widetext}
where, for convenience,  we chose  a mixed representation in terms of $f_n(\mathbf{r},t)$ and $\tilde{f}_k(\mathbf{q},t)$.

%%%%%%%%%%%%%%%%%%%%%
\subsubsection{Point-wise interactions} \label{app:closure}
%%%%%%%%%%%%%%%%%%%%%

We first discuss the limit of point-wise interactions, corresponding to \smash{$I(\mathbf{r})=\delta(\mathbf{r})$ and $\tilde{I}=1$}, which has been widely used in models that contain polar ($k=1$) and nematic ($k=2$) alignment interactions~\cite{bert09,bert15,lieb16} as given in Eq.~(\ref{eq:micrmod}). In this case, Eqs.~(\ref{eq:gensys}) simplify to a spatially homogeneous system of equations
\begin{equation}\label{eq:modhierarch}
\partial_tf_n=\frac{gn}{2}\left(f_{n-k}f_k-f_{n+k}f_{-k}\right)-D_rn^2f_n.
\end{equation}
This system can be further split into a subset of coupled equations for the modes $n=jk$ with integers $j\ne0$:
\begin{equation}\label{eq:katdyn}
\partial_tf_{jk}=\frac{gjk}{2}\left(f_{(j-1)k}f_k-f_{(j+1)k}f_{-k}\right)-D_r(jk)^2f_{jk}
\end{equation}
and a system of equations for the modes $f_n$ with \hbox{$n\ne jk$}, where the latter modes always vanish at long times. We then generalize the standard closure assumption of a fast relaxation of the next coupled mode~\cite{bert09} to the case of a $k$-atic system, which corresponds to assuming $\partial_tf_{2k}=0$ and $f_{sk}=0$ for integers $s\ge3$. From Eqs.~(\ref{eq:katdyn}), we find in this case a steady-state value for $f_{2k}$ and consequently the closed coarse-grained dynamics 
\begin{equation}\label{eq:fkdyn}
\partial_tf_k=\frac{g\rho k}{2}\left(1-\frac{2D_rk}{g\rho}\right)f_k-\frac{g^2}{8D_r}|f_k|^2f_k.
\end{equation}
Using the dimensionless $k$-atic mode $\psi_k=f_k/\rho$ in Eq.~(\ref{eq:fkdyn}), we arrive at the final coarse-graining result given in Eq.~(\ref{eq:kcoarse}).

\subsubsection{$k$-atic order parameter in particle simulations}
To approximate the normalized $k$-atic mode~$\psi_k=f_k/\rho$ in particle simulations, we use the classical $k$-atic order parameter~\cite{chaik00}
\begin{equation}\label{eq:psikdef}
\psi_k(\mathbf{r}_i)=\frac{1}{|\mathcal{N}_i|}\sum_{j\in\mathcal{N}_i}e^{ik\alpha_j},
\end{equation}
where the sum is evaluated with respect to all particles~$j$ within a neighborhood~$\mathcal{N}_i$ of particle~$i$. This can be motivated as follows: Using Eqs.~(\ref{eq:fprobdef}) and (\ref{eq:Fourierang}), and ignoring time for brevity, we have 
\begin{align}
f_k(\mathbf{r}_i)&=\int_0^{2\pi} d\alpha\,\sum_j\langle\delta(\mathbf{r}_i-\mathbf{r}_j)\delta(\alpha-\alpha_j)\rangle\,e^{ik\alpha}\nonumber\\
&\approx\frac{1}{\pi R_i^2}\left\langle\int_0^{2\pi} d\alpha\,\sum_{j\in\mathcal{N}_i}\delta(\alpha-\alpha_j)\,e^{ik\alpha}\right\rangle\nonumber\\
&\approx\rho\langle\psi_k(\mathbf{r}_i)\rangle.
\end{align}
Here, $R_i$ denotes the neighborhood radius and in the last step we have used that particles are homogeneously distributed, such that $|\mathcal{N}_i|/(\pi R_i^2)\approx\rho$. To compute the order parameter $\langle\psi_k(\mathbf{r}_i)\rangle$ for stationary states (Figs.~\ref{fig2}a, \ref{fig2x}a,~\ref{fig4}a, \ref{figS2}, \ref{figS3}), we have replaced the Gaussian white-noise average by temporal averages.

\subsubsection{Linearization with an arbitrary interaction kernel}
For fixed, homogeneous particle number density $\rho=f_0$ the mode coupling terms under the integral in the system of Eqs.~(\ref{eq:gensys}) only contain a linear contribution when \hbox{$|n|=k$}. As a result, the full linearization of Eq.~(\ref{eq:gensys}) around $f_n=0$ is for $|n|\ne k$ simply given by 
\begin{equation}
\partial_t f_n=-D_rn^2f_n,
\end{equation}
and reads for $|n|=k$:
\begin{equation}
\partial_t\tilde{f}_k\left(\mathbf{q},t\right)=\frac{g\rho k}{2}\left[\tilde{I}\left(\mathbf{q}\right)-\frac{2D_rk}{g\rho}\right]\tilde{f}_k\left(\mathbf{q},t\right).\label{eq:disprelq}
\end{equation}
Equation~(\ref{eq:disprelq}) defines the dispersion relation for the $k$-atic mode dynamics near the disordered state and holds for arbitrary spatial interaction kernels. 

%%%%%%%%%%%%%%%%%%%%%
\subsection{Approximation of the pseudo-differential operator and dispersion relation}
\label{appsec:psdiffop}
%%%%%%%%%%%%%%%%%%%%%%
In the microscopic model Eq.~\eqref{eq:micrmod}, we consider an isotropic interaction neighborhood, such that the interaction kernel in Fourier-space must be an even function that depends only on the wave vector amplitude $q=|\mathbf{q}|$. Assuming that $\tilde{I}(x)$ can be expanded in a suitable power-series
\begin{equation}\label{eq:ps}
\tilde{I}(x)=1+\sum_{j=1}^{\infty}\beta_jx^{2j},
\end{equation}
Eq.~(\ref{eq:disprelq}) has an equivalent interpretation in real space that is given by
\begin{equation}
\partial_tf_k\left(\mathbf{r},t\right)=\frac{g\rho k}{2}\left[\tilde{I}\left(R_{\alpha}^2\nabla^2\right)-\frac{2D_rk}{g\rho}\right]f_k\left(\mathbf{r},t\right),\label{eq:disprel}
\end{equation}
where $\tilde{I}(R_{\alpha}^2\nabla^2)$ represents a pseudo-differential operator that is defined by the power series Eq.~(\ref{eq:ps}).

\begin{figure}%[t]
\centering
	\includegraphics[width = 1\columnwidth]{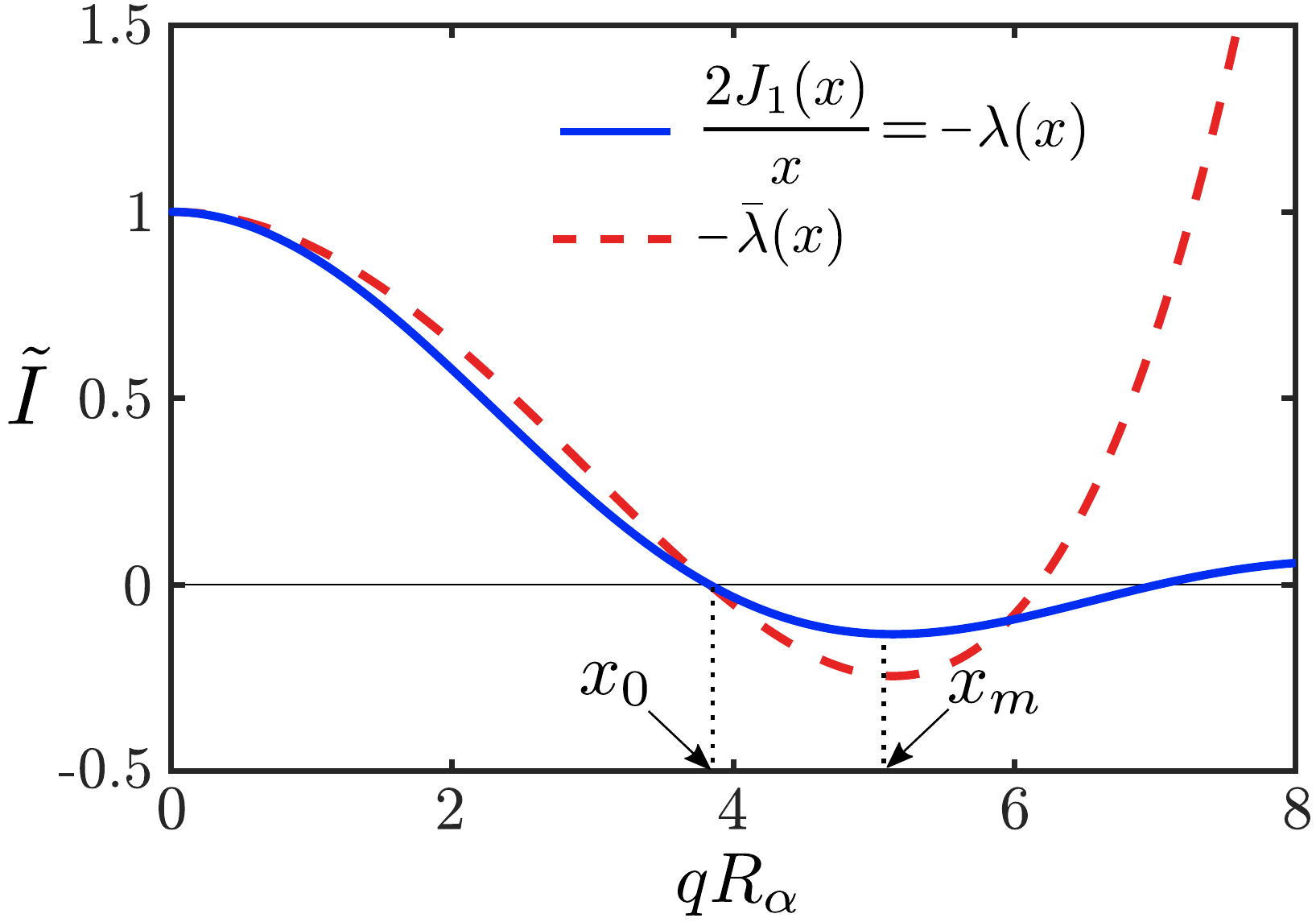}
	\caption{Fourier-space representation of the interaction kernel $\tilde{I}(x)=-2J_1(x)/x$ (solid line) used in the microscopic particle model Eq.~\eqref{eq:micrmod}, where J$_1$ denotes the Bessel function of the first kind. The linearized mode dynamics Eq.~(\ref{eq:disprelq}) implies for $D_r/(|g|\rho)\ll1$ that \smash{$\lambda=-\tilde{I}(q)$} essentially represents the dispersion relation of the $k$-atic particle model with anti-aligning interactions ($g<0$). An empirical approximation $-\bar{\lambda}=1-\beta_1x^2+\beta_2x^4$ (dashed line) is then chosen such that the smallest unstable wavelength set by the first root $x_0$ and the most unstable wavelength set by $x_m$ are the same for $\bar{\lambda}$~and~$\lambda$. This yields Eq.~(\ref{eq:LinDynGen}) with $\beta_1\approx0.1$ and $\beta_2\approx0.002$.}
	\label{figS1}
\end{figure}

\begin{figure}[!h]
\centering
	\includegraphics[width = 1\columnwidth]{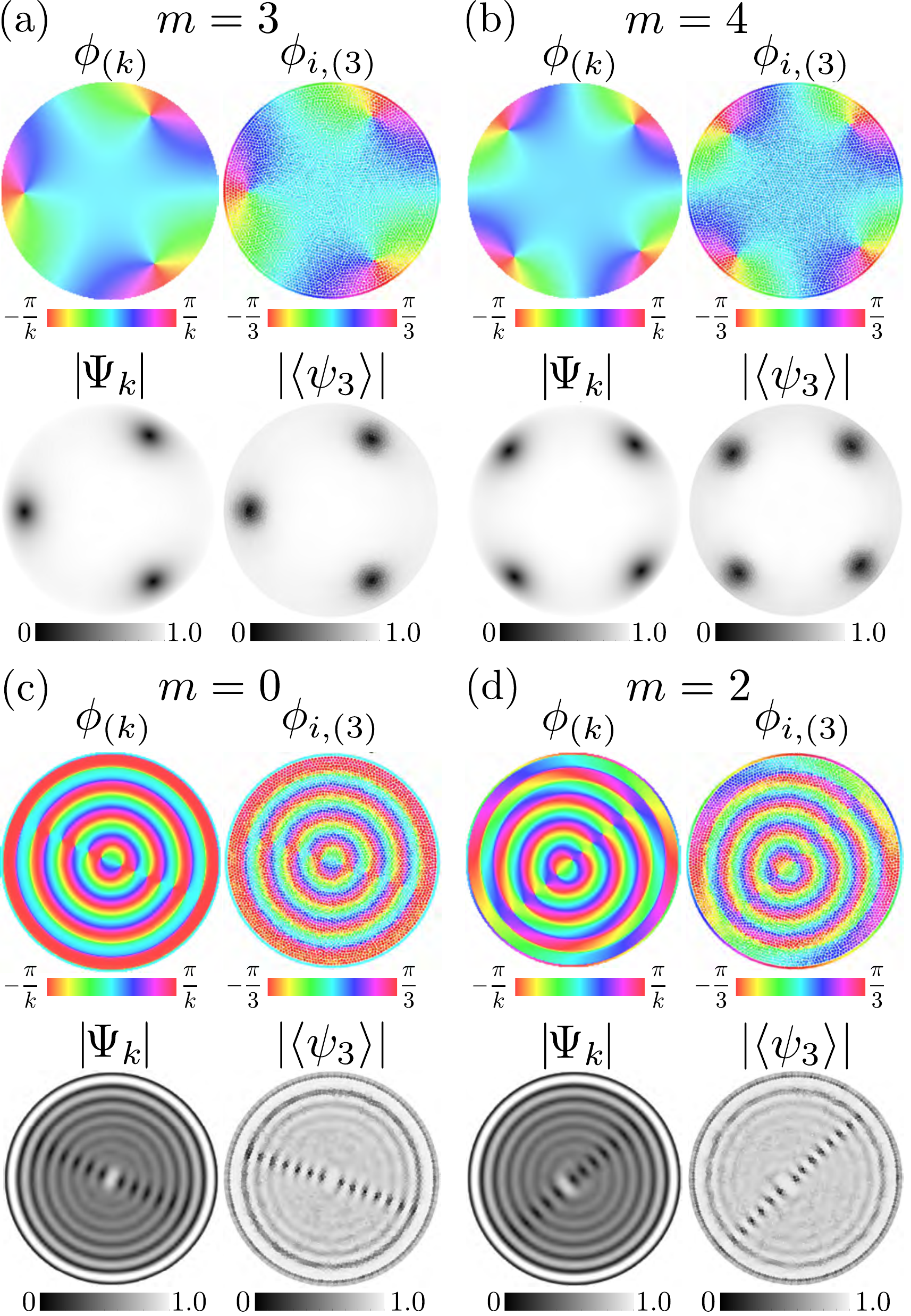}
	\caption{Additional stationary and long-lived solutions of the \lq real\rq\ GL equation and particle model Eq.~(\ref{eq:micrmod}) on a unit disk. (a)~Stationary solutions emerging from random initial conditions. Boundary anchorings Eq.~(\ref{eq:bc1}) (left, Eq.~(\ref{eq:RGLgen}) with $\mathcal{L}=L^2\nabla^2$) and Eq.~(\ref{eq:bcpm}) (right, aligning particle dynamics Eq.~(\ref{eq:micrmod}) with $k=3$) for $m=3$ and boundary anchoring profile $\gamma_k=\theta$ were used. All other parameters as in Fig.~\ref{fig2}c,d. (b)~Same as (a) for boundary anchoring with~$m=4$. (c)~Stationary solutions of the complex SH equation emerging from random initial conditions (left) and long-lived solutions in the particle model with anti-aligning interactions (right) for boundary anchorings with $m=0$. All other parameters as in Fig.~\ref{fig4}a. Initial director orientation in particle simulations were sampled from stationary solution of the complex SH equation and evolved according to Eq.~(\ref{eq:micrmod}) until $D_rt=500$. (d)~Same as (c) for boundary anchoring with~$m=2$.}
\label{figS2}
\end{figure}

The equally weighted summation over orientational interactions with particles in an isotropic neighborhood of radius $R_{\alpha}$ corresponds to an interaction kernel $\hat{I}(r)=\Theta(r-R_{\alpha})$ in Eq.~(\ref{eq:cg1}). The Fourier-transform of the appropriately normalized kernel $I(r)=\hat{I}(r)/(2\pi R_{\alpha}^2)$ defined by Eq.~(\ref{eq:Ir}) then reads
\begin{equation}\label{eq:Jkern}
\tilde{I}(x)=\frac{2J_1(x)}{x},
\end{equation}  
where $J_1$ denotes the Bessel function of the first kind~(Fig.~\ref{figS1}). 

%%%%%%%%%%%%%%%%%%
%%%%%%%%%%%%%%%%%%
%%%%%%%%%%%%%%%%%%
%%%%%%%%%%%%%%%%%%

To map the coarse-grained mode dynamics Eq.~(\ref{eq:disprelq}) for $k$-atic alignment interactions ($g>0$) to the GL as a mean-field model, we consider the Taylor series of \smash{$\tilde{I}(x)$} given in Eq.~(\ref{eq:Jkern}) around $x=0$, which implies a non-vanishing coefficient $\beta_1=-1/8$ in Eq.~(\ref{eq:ps}) and is accurate to $\mathcal{O}(x^4)$. Using this expansion for the operator in Eq.~(\ref{eq:disprel}) leads to Eq.~(\ref{eq:kcoarseany}) and allows for the identification of an effective mean-field correlation length $L$ in terms of the microscopic interaction radius~$R_{\alpha}$ [Eq.~(\ref{eq:Ldef})].

\par
To connect Eq.~(\ref{eq:disprelq}) to the complex SH equation as a mean-field description of anti-aligning $k$-atics ($g<0$), we first note that, for $|g|\rho\gg1$, we can identify \hbox{$\lambda(q)=-\tilde{I}(q)$} as the dispersion relation that describes the stability ($\lambda<0$) or instability ($\lambda>0$) of homogeneous states under perturbations with wavelength $2\pi/q$. Consequently, the first interval in which the Fourier-space representation of the interaction kernel (Fig.~\ref{figS1}, solid line) changes its sign indicates a band of unstable wave vectors. Because this sign-change is not captured by the Taylor series of $\tilde{I}(x)$ up to fourth order around $x=0$, we instead empirically define an approximation $\bar{\lambda}$ such that $i)$~$\bar{\lambda}(q)={\lambda}(-q)$ and $\bar{\lambda}(0)=1$, and $ii)$~the smallest unstable wave-vector and the most unstable wave-vector are approximately the same for $\bar{\lambda}(q)$ and $\lambda(q)$~(Fig.~\ref{figS1}, dashed line). With this approximation, Eq.~(\ref{eq:disprelq}) implies the real-space representation Eq.~(\ref{eq:LinDynGen}) for the linearized mode dynamics and a length-scale matching as given~in~Eq.~(\ref{eq:L1L2def}).

Finally, we note that a generalization of the hierarchy of mode Eqs.~(\ref{eq:modhierarch}) to the case of finite-range interaction kernels still allows to decouple the dynamics of modes $n=jk$ for arbitrary integer $j$ from all other modes with~\hbox{$n\ne jk$}. However, the closure assumptions described in Appendix~\ref{app:closure} lead in this case to additional nonlinear terms in the final dynamic equation of the mode $f_k$. These terms, which are $\mathcal{O}(f_k^2\nabla^2f_k)$ to lowest order, have for simplicity been neglected in Eqs.~(\ref{eq:kcoarseany}) and (\ref{eq:LinDynGen}).

%%%%%%%%%%%%%%%%%%%%%%%%%%%%
\section{Landau-de Gennes (LdG) theory of DLC\lowercase{s} with \lowercase{$k$}-fold symmetry}
\label{app:NematicLCs}
%%%%%%%%%%%%%%%%%%%%%%%%%%%%

We first explain how the LdG theory of nematic liquid crystals can be naturally generalized to describe DLCs with arbitrary $k$-fold symmetries. Subsequently, we will formally map the resulting relaxation dynamics onto the mean-field Eq.~\eqref{eq:RGLgen} with $\mathcal{L}=L^2\nabla^2$ and discuss simple fractional defect solutions in free space. The SH mean-field theory can be discussed in an analog fashion~\cite{oza16}.

%%%%%%%%%%%%%%%%%%%%%
\subsection{Free energy of $k$-atic DLCs in 2D}\label{app:freeEnLCs}
%%%%%%%%%%%%%%%%%%%%%

Properties of DLCs can be conveniently studied using appropriate free-energy functionals. Typically, these functionals are constructed using vector and tensor-valued fields to encode the underlying microscopic symmetries. Classic examples of this approach are the Frank free energy~\cite{chand92} for polar liquid crystals ($k=1$) and LdG free energy~\cite{dege93} for nematics ($k=2$) that are formulated in terms of vectors and traceless symmetric (nematic) tensors, respectively. A particular advantage of this formulation is that a free energy can be systematically constructed as an expansion in terms of the available scalar (rotationally invariant) contractions that can be formed by the given tensorial objects.

%%%%%%%%%%%%%%%%%%
%%%%%%%%%%%%%%%%%%
%%%%%%%%%%%%%%%%%%
%%%%%%%%%%%%%%%%%%

\par
A generalization of this approach to arbitrary $k$-fold symmetric systems can be realized using traceless symmetric tensors of rank~$k$, denoted by \smash{$Q_{i_1...i_k}^{(k)}$}, which are invariant under rotations of $2\pi/k$. $k$-atic tensors in two dimensions have only two independent degrees of freedom for any $k$. This is specific to two dimensions and the total number of independent degrees of freedom increases with $k$ in any higher dimension. In 2D, it is convenient to choose one of these degrees of freedom as
\begin{equation}\label{eq:rotinv}
Q_{(k)}=\sqrt{2^{1-k}\,Q^{(k)}_{i_1...i_k}Q^{(k)}_{i_1...i_k}}\ ,
\end{equation}  
which can be identified as the local $k$-atic order. The second degree of freedom can then be chosen as the local orientation of the $k$-atic director $\phi_{(k)}\in\left(-\frac{\pi}{k},\frac{\pi}{k}\right]$. Adopting this parametrization, $k$-atic tensors \smash{$Q^{(k)}_{i_1...i_k}$} are uniquely determined by specifying the two tensor components
\begin{subequations}
\label{eq:Q_comp}
\begin{align}
\label{eq:Q1}
a_k&:=Q^{(k)}_{x...xx}=Q_{(k)}\cos k\phi_{(k)},\\
b_k&:=Q^{(k)}_{x...xy}=Q_{(k)}\sin k\phi_{(k)},
\label{eq:Q2}
\end{align}
\end{subequations}
with all other components being implied by the index symmetry and tracelessness. Importantly, the tensor parametrization given in Eqs.~(\ref{eq:Q_comp}) can be used to define the complex $k$-atic order parameter
\begin{equation}\label{eq:ordtens}
\Psi_k=a_k+ib_k.
\end{equation}
This definition is equivalent to $\Psi_k$ given in Eq.~(\ref{eq:psik}) and explicitly relates the magnitude $|\Psi_k|=Q_{(k)}$ of $k$-atic order and the $k$-atic phase~$\phi_{(k)}$ as introduced in Eq.~(\ref{eq:phasedef}) to a representation of $k$-atic DLCs in terms of traceless symmetric tensors of rank $k$.

\par
To connect a mean-field description of $k$-atic systems in terms of such $k$-atic tensors to Eq.~\eqref{eq:RGLgen}, we start from the generic free energy  
\begin{equation}
F_k^{\text{GL}}=\int d^2r\left[f_h+\frac{L^2}{2}\left(\partial_jQ^{(k)}_{i_1,...,i_k}\right)^2\right],\label{eq:freeen}
\end{equation}
where $L$ is a parameter describing the $k$-atic system, and homogeneous contributions $f_h$ must consist of rotational invariants that can be formed by $k$-atic tensors. In Eq.~(\ref{eq:freeen}) the role of $L$ as an effective length scale that penalizes bending of the local $k$-atic director field becomes explicit. 

To determine a minimal form of the function $f_h$ in Eq.~(\ref{eq:freeen}) that is allowed by the underlying symmetries, one has to analyze the possible contractions between \hbox{$k$-atic} tensors that can be constructed to form rotational invariants (scalars). Using the properties of general $k$-atic tensors, one can show that cubic contractions between $k$-atic tensors must vanish for arbitrary $k$ -- a fact that is well-known for nematics $k=2$. Hence, a minimal LdG expansion of $f_h$ in the free energy Eq.~\eqref{eq:freeen} is, for any $k$, given~by
\begin{equation}
f_h=\frac{A}{2}Q^{(k)}_{i_1...i_k}Q^{(k)}_{i_1...i_k}+\frac{B}{2^{k+1}}\left(Q^{(k)}_{i_1...i_k}Q^{(k)}_{i_1...i_k}\right)^2,
\end{equation}
where $A\in\mathbb{R}$ and $B>0$ are constant material parameters. The relaxation dynamics \smash{$\tau\partial_tQ_{i_1...i_k}^{(k)}=-\delta F_{k}^{\text{GL}}/\delta Q_{i_1...i_k}^{(k)}$} thus takes the form
\begin{equation}
\hspace{-0.1cm}\tau\partial_t Q^{(k)}_{i_1...i_k}=-\left(A+BQ_{(k)}^2\right)Q^{(k)}_{i_1...i_k}+ L^2\nabla^2Q^{(k)}_{i_1...i_k},\label{eq:Qrelax}
\end{equation}
where $Q_{(k)}$ is defined in Eq.~(\ref{eq:rotinv}). Using Eqs.~(\ref{eq:Q_comp})~and~(\ref{eq:ordtens}), we see that the relaxation dynamics Eq.~(\ref{eq:Qrelax}) is indeed equivalent to mean-field theory of aligning $k$-atic particles, Eq.~(\ref{eq:RGLgen}) with $\mathcal{L}=L^2\nabla^2$, corresponding to a \lq real\rq\ GL equation for a complex order parameter $\Psi_k$. 

The latter equivalence becomes also evident on an energetic level through the energy functional $\mathcal{E}_k$ given in Eq.~(\ref{eq:ComplexGLE}): For $F_k^{\text{GL}}$ given in Eq.~(\ref{eq:freeen}), Eqs.~(\ref{eq:Q_comp})~and~(\ref{eq:ordtens}) imply \hbox{$\mathcal{E}_k=2^{2-k}F_k^{\text{GL}}$} and the relaxation dynamics $\tau\partial_t\Psi_k=-\delta\mathcal{E}_k/\delta\Psi_k^*$ yields the same \lq real\rq\ GL equation for the complex order parameter~$\Psi_k$ that we have just identified as being equivalent to~Eq.~(\ref{eq:Qrelax}).

%%%%%%%%%%%%%%%%%%%%%%%%%%%%%%
\subsection{Fractional point defect solutions in free space}
\label{appsec:ptdef}
%%%%%%%%%%%%%%%%%%%%%%%%%%%%%%
The simplest scenario to study fractional defects in the GL equation, or equivalently in Eq.~(\ref{eq:Qrelax}), is to consider a limit $B=-A\rightarrow\infty$, such that \hbox{$|\Psi_k|=\sqrt{-A/B}=1$} and the system resides in a perfectly ordered state. Stationary solutions are then determined by
\begin{equation}
\nabla^2\phi_{(k)}=0.
\label{eq:LapPsi}
\end{equation} 
The regularity of the complex order parameter $\Psi_k$ away from the defect demands \hbox{$k[\phi_{(k)}\left(r,\theta+2\pi\right)-\phi_{(k)}\left(r,\theta\right)]=2\pi m$} for any integer~$m$, where $(r,\theta)$ denote cylindrical coordinates. Hence, physically permissible topological defect solutions of Eq.~(\ref{eq:LapPsi}) can be written as
\begin{equation}\label{eq:phikdefsol}
\phi_{(k),m}(\theta)=\frac{1}{k}\text{arg}\left(e^{im\theta}\right),
\end{equation}
which provides an example for a fractional defect state with topological charge $m/k$ as defined by Eq.~(\ref{eq:defch}).
\par
For finite values of $A$ and $B$, the scaling behavior close to and far away from $m/k$-defects can be obtained by following the approach of Ref.~\cite{bode88}. Using an ansatz $\Psi_k=Q_0(r)\exp[ik\phi_{(k)}(\theta)]$ in the GL, one again finds stationary fractional $m/k$-defect solutions Eq.~(\ref{eq:phikdefsol}), where the magnitude $Q_0(r)$ is now a function of the distance~$r$ from the defect center. In the $k$-atically ordered regime $A<0$, the magnitude increases near the defect ($r\ll L$)  as \mbox{$Q_0\sim (r/L)^m$} and converges far away from the defect ($r \gg L$) to the value $\sqrt{-A/B}$ with an asymptotic scaling behavior of \mbox{$Q_0\sim1-(mL)^2/(2|A|r^2)$}.

%%%%%%%%%%%%%%%%%%%
\section{Landau-Bazovskii energy and Swift-Hohenberg equation}
\label{app:SH}
%%%%%%%%%%%%%%%%%%%
In this appendix, we introduce an effective energy that governs the mean-field dynamics of $k$-atic particles with anti-aligning interactions, the Landau-Brazovskii energy~\cite{braz75}, and use it to identify boundary conditions for numerical simulations. We then derive analytic stationary solutions that were used to explain the wavelength-doubling and the chiral symmetry breaking, and present additional examples that demonstrate the close agreement between this theory and the particle~model. 

\subsection{Free energy and boundary~conditions}\label{app:freeEnSH}
The mean-field theory of particles with anti-aligning interactions can be written as relaxation dynamics $\tau\partial_t\Psi_k=-\delta\mathcal{E}^{\text{LB}}_k/\delta\Psi_k^*$ with energy
\begin{align}\label{eq:ComplexSHE}
\mathcal{E}_k^{\text{LB}}&=\int d^2r\left(A\left|\Psi_k\right|^2+\frac{B}{2}\left|\Psi_k\right|^4\right.\nonumber\\
&\hspace{1.7cm}\left.-L_1^2\left|\nabla\Psi_k\right|^2+L_2^4\left|\nabla^2\Psi_k\right|^2\right),
\end{align}
where $\left|\nabla^2\Psi_k\right|^2=\left(\nabla^2\Psi_k\right)\left(\nabla^2\Psi_k^*\right)$ and $\mathcal{E}_k^{\text{LB}}$ is known as Landau–Brazovskii energy~\cite{braz75}. A general variation of Eq.~(\ref{eq:ComplexSHE}) with respect to $\Psi^*$ yields
\begin{widetext}
\begin{equation}\label{eq:varESg}
\delta\mathcal{E}_k^{\text{LB}}=\int_Sd^2r\delta\Psi_k^*\left(A+B\left|\Psi_k\right|^2+L^2_1\nabla^2+L^2_2\nabla^2\nabla^2\right)\Psi_k
+\int_{\partial S}ds\mathbf{n}\cdot\left[\delta\Psi_k^*\left(L_1^2\nabla\Psi_k-L_2^2\nabla\nabla^2\Psi_k\right)+L_2^2(\nabla^2\Psi_k)\nabla\delta\Psi_k^*\right],
\end{equation}
\end{widetext}
where the second term denotes a line integral with boundary normal $\mathbf{n}$ that collects all boundary terms arising from the variation. For the equilibrium condition \hbox{$\delta\mathcal{E}_k^{\text{LB}}=0$}, the first integral in Eq.~(\ref{eq:varESg}) implies the complex-valued Swift-Hohenberg equation discussed in the main text Sec.~\ref{sec:SH} (Eq.~(\ref{eq:RGLgen}) with $\mathcal{L}=-L_1^2\nabla^2-L_2^2\nabla^2\nabla^2$). From the condition of vanishing boundary terms in Eq.~(\ref{eq:varESg}), suitable boundary conditions can be derived. In particular, we have fixed the order parameter at the boundary through specific anchoring profiles ($\Rightarrow\delta\Psi_k=0$) and additionally imposed $\nabla^2\Psi_k|_{\partial S}=0$ in numerical simulations (see Appendix~\ref{app:NumMeth}). Therefore, stationary solutions of Eq.~(\ref{eq:RGLgen}) fulfill $\delta\mathcal{E}_k^{\text{LB}}=0$ and extremize the energy given in Eq.~(\ref{eq:ComplexSHE}).

\subsection{Analytic solutions}\label{app:ansolSH}
We derive an analytic solution of $\delta\mathcal{E}_k^{\text{LB}}/\delta\Psi_k^*=0$ near the critical transition at $A=A_*$ (see Fig.~\ref{fig4}c). This solution recapitulates the stationary patterns shown in Fig.~\ref{fig4}, including the wavelength-doubling between phase and amplitude patterns and the emergence of chiral texture patterns when moving away from the critical point $A_*$ (Fig.~\ref{figtheo}).

For this derivation, we neglect quartic terms $\sim|\Psi_k|^4$ in Eq.~(\ref{eq:ComplexSHE}) and seek complex order parameter fields $\Psi_k$ that solve
\begin{equation}\label{eq:linSH}
A\Psi_k+L_1^2\nabla^2\Psi_k+L_2^4\nabla^2\nabla^2\Psi_k=0.
\end{equation}
We write $A=A_*-\Delta A$, where $A_*=q_0^4L_2^4$ is the critical value of the linear instability and $q_0^2=L^2_1/(2L_2^4)$ is the first unstable wavenumber; see discussion below Eq.~(\ref{eq:L1L2def}). With these definitions, Eq.~(\ref{eq:linSH}) can via a square completion be cast into the form
\begin{equation}\label{eq:biHelm}
\left(\nabla^2+q_+^2\right)\left(\nabla^2+q_-^2\right)\Psi_k=0,
\end{equation}
where we have defined $q_{\pm}^2=q_0^2\pm\sqrt{\Delta A}/L_2^2$. Equation~(\ref{eq:biHelm}) represents a bi-Helmholtz equation~\cite{askham18} that can be solved in polar coordinates $(r,\theta)$ by
\begin{equation}\label{eq:complSHsol}
\Psi_k(r,\theta)=\sum_{m=0}^{\infty}\left[\mu_mJ_m(q_-r)+\nu_mJ_m(q_+r)\right]e^{im\theta},
\end{equation}
where $J_m(x)$ are Bessel functions of the first kind, and $\mu_m$ and $\nu_m$ are possibly complex integration constants.

\subsubsection{Wavelength-doubling between amplitude \\ and phase patterns}\label{app:wldiff}

At the critical point $A=A_*$, we have \hbox{$q_-=q_+=q_0$}, and solutions Eq.~(\ref{eq:complSHsol}) will be of the form $\Psi_k\sim J_m(q_0r)e^{im\theta}$, consistent with the boundary anchoring Eq.~(\ref{eq:bc1}) for $\gamma_k=\theta$. Consequently, solutions of this kind contain a topological defect of charge $q_\text{d}=m/k$ at $r=0$ and they recapitulate the factor 2 difference in the wavelength of amplitude patterns $|\Psi_k|$ and phase patterns $\phi_{(k)}$ seen in Fig.~\ref{fig4}. To illustrate this for $m=0$ (Figs.~\ref{fig4}a and~\ref{figtheo}a), we note that the solution $\Psi_k\sim J_0(q_0r)$ corresponds to
\begin{subequations}
\label{eq:J0sol}
\begin{align}
|\Psi_k|&\sim|J_0(q_0r)|\\
\phi_{(k)}&=\left\{
\begin{matrix*}[l]
0\text{\hspace{1cm}for }J_0(q_0r)\ge0\\
\pi/k\text{\hspace{0.6cm}for }J_0(q_0r)<0.
\end{matrix*}
\right.
\end{align}
\end{subequations}
As $J_0(q_0r)$ oscillates with wavelength $\lambda_0\sim 2\pi/q_0$ around zero, Eqs.~(\ref{eq:J0sol}) imply that phase pattern also have wavelength $\lambda_0$, while the amplitude pattern wavelength is $\lambda_0/2$, precisely as observed in numerical simulations~(Fig.~5a). Similarly, for $m=1$ (Fig.~\ref{figtheo}b), the solution $\Psi_k\sim J_1(q_0r)e^{i\theta}$ corresponds to
\begin{align*}
|\Psi_k|&\sim|J_1(q_0r)|\\
\phi_{(k)}&=\left\{
\begin{matrix*}[l]
\arg(e^{i\theta})/k\text{\hspace{0.9cm}for }J_1(q_0r)\ge0\\
\arg(e^{i(\theta-\pi)})/k\text{\hspace{0.3cm}for }J_1(q_0r)<0
\end{matrix*}
\right.,
\end{align*}
which recapitulates the amplitude and phase patterns shown in the right-most panel of Fig.~\ref{fig4}b.

\subsubsection{Analytic solutions with chiral texture patterns}\label{app:chirsol}
We derive sufficient conditions for the emergence of chiral texture patterns in the mode $m=1$ of the analytic solution Eq.~(\ref{eq:complSHsol}), denoted in the following~as
\begin{equation}\label{eq:psi1sol}
\Psi^{(1)}_k=\left(\mu_1J_-+\nu_1J_+\right)e^{i\theta}
\end{equation}
with $J_{\pm}:=J_1(q_{\pm}r)$. Note, that for any choice of complex integration constants $\mu_1$ and $\nu_1$ in Eq.~(\ref{eq:psi1sol}) the amplitude \smash{$|\Psi^{(1)}_k|$} is independent of the polar angle~$\theta$ and therefore remains azimuthally symmetric. This is consistent with the various amplitude patterns shown in Fig.~\ref{fig4} of the main text. To determine for which parameters the phase pattern $\phi_{(k)}$ of \smash{$\Psi^{(1)}_k$} could be chiral, we note that for any complex field~$\Psi=|\Psi|e^{i\phi}$, gradients of the phase~$\phi$ can be conveniently computed from
\begin{equation}\label{eq:phasgradid}
\nabla\phi=|\Psi|^{-2}\text{Im}\left(\Psi^*\nabla\Psi\right).
\end{equation}
Chirality in texture patterns around defects at $r=0$ can be detected if the radial part of this gradient, $\mathbf{e}_r\cdot\nabla\phi$, is different from zero with a fixed sign across the domain, which inspired the phase chirality parameter given in Eq.~(\ref{eq:chphasepara}). From Eq.~(\ref{eq:phasgradid}), we find for $\Psi^{(1)}_k$ and $A\le A_*$ the expression
\begin{equation}\label{eq:dphi}
\mathbf{e}_r\cdot\nabla\phi_{(k)}=\frac{\text{Im}(\mu_1\nu_1^*)}{k|\Psi^{(1)}_k|^2}\left(J_+\partial_rJ_--J_-\partial_rJ_+\right).
\end{equation}
From $\mathbf{e}_r\cdot\nabla\phi_{(k)}\ne0$, we find two necessary conditions from Eq.~(\ref{eq:dphi}) for \smash{$\Psi^{(1)}_k$} given in Eq.~(\ref{eq:psi1sol}) to describe chiral texture patterns: (1)~$\mu_1\not\propto\nu_1$, such that at least one of the two integration constants must be complex, and (2)~$q_+\ne q_-\Rightarrow A<A_*$, meaning that the system has to be in a regime in which a finite band of wavenumbers are linearly unstable. The numerical observation that chiral patterns are absent for $A>A_*$, additionally constrains the integration constants to $|\mu_1|=|\nu_1|$.

%%%%%%%%%%%%%%%%%%
%%%%%%%%%%%%%%%%%%
\begin{figure}[!t]
\centering
	\includegraphics[width = 1\columnwidth]{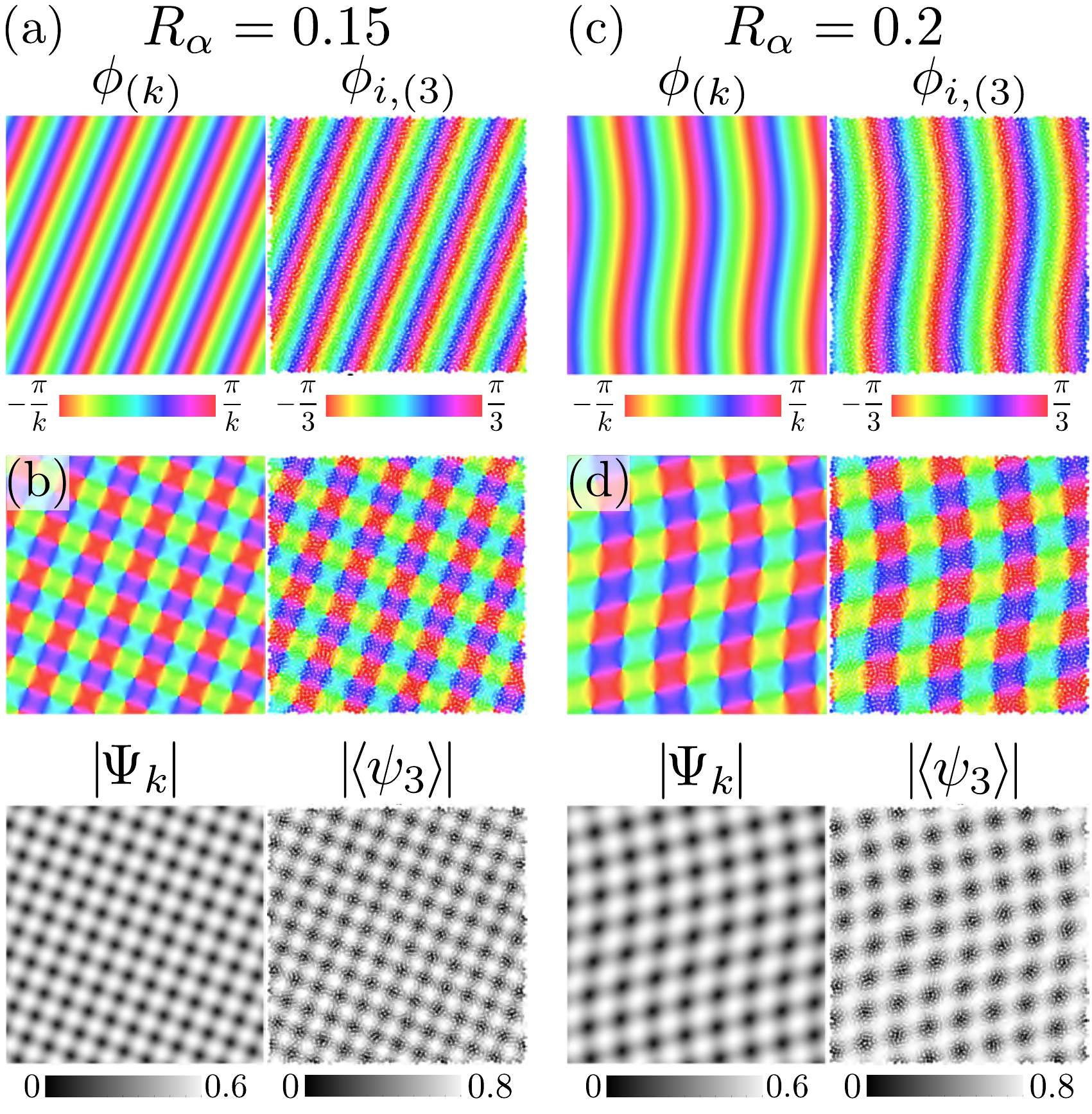}
	\caption{Spontaneous pattern formation in the SH Eq.~(\ref{eq:RGLgen}) with \hbox{$\mathcal{L}=-L_1^2 \nabla^2 - L_2^4 (\nabla^2)^2$} and in the particle model with anti-alignment interactions [Eq.~(\ref{eq:micrmod}) with $g<0$; 5,000~particles] on a unit square with periodic boundaries. (a)~Example of a defect-free wave-like texture pattern that can be found in both models. (b)~Example of a checkerboard-like texture pattern (same color code as in~a) that contains a $\pm 1/k$-defect-lattice visible in the magnitude of the $k$-atic order parameter $|\Psi_k|$ and in the temporal average $\langle\cdot\rangle$ of the magnitude of the microscopic 3-atic order parameter $\psi_3=\sum_{j\in\mathcal{N}_i}\exp(3i\alpha_j)/\left|\mathcal{N}_i\right|$. (c,d)~Similar patterns for a different microscopic interaction radius and an accordingly modified SH equation. The phase patterns are reminiscent of those found in  polar models of pinwheels  in cortical visual maps, see Fig.~4 in~\cite{lee2003}. Parameters: $A=1$, $B=1$, $L_1=0.305R_{\alpha}$ and $L_2=0.205R_{\alpha}$ (SH equation) and $k=3$, $g=-1$, and $D_r=1$ (particle model).}
\label{figS3}
\end{figure}
%%%%%%%%%%%%%%%%%%
%%%%%%%%%%%%%%%%%%

\subsection{Patterns on periodic domains}
For completeness, we have additionally studied the anti-aligning $k$-atic particle model and the complex SH equation on a periodic unit square using numerical simulations (Fig.~\ref{figS3}). In both models and for the parameter regime studied in this work, many different texture patterns spontaneously form. These patterns can be broadly grouped in defect-free wave-like (Fig.~\ref{figS3}a,c) and checkerboard-like patterns~(Fig.~\ref{figS3}b,d). The former have spatially constant order $|\Psi_k|>0$ and are near the critical value $A_*=L_1^4/(4L_2^4)$ essentially given by plane phase-waves of the form $\Psi_k\sim e^{i\mathbf{q}_0\cdot\mathbf{r}}$ with $q_0^2=L_1^2/(2L_2^4)$. The latter represent $\pm 1/k$-defect-lattices, as clearly visible in the order parameter magnitudes, with vanishing total topological charge. We note that defect-free patterns generally have a lower energy than checkerboard patterns. However, this energy difference is small and inhomogeneous patterns are generally preferred by the dispersion relation $\lambda(q)=-\tilde{I}(q)$ (see Fig.~\ref{figS1}, which implies that $q=0$ is stable); this is sufficient for checkerboard patterns to emerge frequently as final stationary textures when starting simulations with random initial conditions. Finally, we found that stationary texture patterns that occur in the complex SH equation are -- for all cases that were tested -- also stationary in the particle model, if the microscopic directors are initialized with the corresponding $k$-atic phase fields. The opposite is not true: The particle dynamics sometimes gets stuck in long-lived irregular patterns that are not stationary when used in the SH equation.

\section{Numerical simulations}
\label{app:NumMeth}
In the following, we we discuss the coefficient matching and comparison of relaxation time scales between mean-field model and microscopic simulations. Finally, we summarize details of the different methods and work-flows that have been used to generate the numerical results presented in this work. Furthermore, 

\subsection{Matching homogeneous mean-field coefficients with parameters of the microscopic model}
To match the homogeneous coefficients $A$ and $B$ in Eq.~(\ref{eq:RGLgen}) to the microscopic dynamics, we note that the latter was considered in a regime of high particle density, viz. $D_r/(|g|\rho)\ll1$. In this case, Eq.~(\ref{eq:abar}) implies $\bar{A}\approx-1$ for aligning interactions ($g>0$) and $\bar{A}\approx1$ for anti-aligning interactions ($g<0$). Accordingly, we have throughout this work set $A=-1$ for simulations of the GL equation [Eq.~(\ref{eq:RGLgen}) with $\mathcal{L}=L^2\nabla^2$] and $A=1$ for simulations of the SH equation [Eq.~(\ref{eq:RGLgen}) with $\mathcal{L}=-L_1^2 \nabla^2 - L_2^4 (\nabla^2)^2$]. The coefficient $B$ was set empirically: Away from defects, we expect $|\Psi_k|\simeq|\psi_k|\approx1$ for an ordered state in the microscopic model, which is ensured in the GL equation by setting $B=1$. The same value is adopted in simulations of the SH equation, where it also leads to good agreement with order parameter magnitudes  of the microscopic model with $g<0$.

\subsection{Comparison of characteristic time scales}
The relaxation time scale introduced with the generalized GL Eq.~(\ref{eq:RGLgen}) is given by $\tau$. From the coarse-graining result Eq.~(\ref{eq:kcoarse}), we expect $\tau$ to be comparable to the time scale $\bar{\tau}=2/(|g|k\rho)$ if the mean-field parameters and operators $\mathcal{L}$ are matched according to Eqs.~(\ref{eq:abar}), (\ref{eq:Ldef}) and (\ref{eq:L1L2def}). To test this, we have to compare observations from a dynamic process as described by the generalized GL Eq.~(\ref{eq:RGLgen}) and by the microscopic model Eq.~(\ref{eq:micrmod}). To this end, we refer to the defect relaxation dynamics depicted in Fig.~\ref{fig2}c,d, where the time points of snapshots of the \lq real\rq\ GL equation simulation in units of~$\tau$ are provided in the caption. The time scale of particle simulations was set for practical reasons by the inverse rotational diffusion constant $1/D_r$, i.e. in units of~$\bar{\tau}$ it was given by~$s\bar{\tau}$ with 
\begin{equation}\label{eq:tsfrac}
s=\frac{2D_r}{|g|k\rho}.
\end{equation}
Scaling numerical time points of particle simulations for the given parameters accordingly ($g=0.25$, $k=3$, \hbox{$\rho=4000/\pi$}, $D_r=1$) then leads to the temporal coordinates $t/\bar{\tau}$ of the particle model snapshots listed in the caption of Fig.~\ref{fig2}d. These snapshots were chosen such that they best resemble textures from the mean-field model. From the relative values of corresponding time points, we can estimate $\bar{\tau}/\tau\approx0.4$, indicating that the coarse-graining predicts a slightly faster relaxation dynamics than actually exhibited by the matched mean-field model Eq.~(\ref{eq:RGLgen}).

\subsection{Numerical methods: Mean-field simulations}
Real and imaginary part of the GL Eq.~(\ref{eq:RGLgen}) with \hbox{$\mathcal{L}=L^2\nabla^2$} were simulated separately on the unit disk using the finite element partial differential equation solver provided by Matlab~\cite{matl19}. Boundary anchoring profiles described in Eq.~(\ref{eq:bc1}) were imposed as Dirichlet boundary conditions. 
\par
For the case of the SH equation, Eq.~(\ref{eq:RGLgen}) with \hbox{$\mathcal{L}=-L_1^2 \nabla^2 - L_2^4 (\nabla^2)^2$} was rewritten as a system of two pairs of second order differential equations and $\nabla^2\Psi_k|_{\partial S}=0$ was included as an additional boundary condition (see Appendix~\ref{app:freeEnSH}). The open-source Dedalus framework~\cite{burn20} was used to spectrally solve the SH equation on a periodic domain with $256\times256$ grid points using a Fourier-basis and integration time steps of $dt=10^{-2}$ in units of $\tau$.

\subsection{Numerical methods: Microscopic model}
\label{appsec:partsim}
Bulk particles were first randomly positioned on the respective domain (a unit disk or a square with periodic boundary). For simulations on the unit disk, a single line of boundary particles with fixed positions was additionally placed along the outline of the disk. In the next step, the bulk particles were left to distribute themselves homogeneously in space in the absence of noise via a pair-wise repulsive force $\mathbf{f}\sim\nabla\exp(-r_p^2/L_f^2)$, where $L_f^2=A/N$ for~$N$ particles distributed on a domain of area $A$. After that, all particle positions were kept fixed and Eq.~(\ref{eq:micrmod}) was integrated using the Euler-Maruyama method~\cite{kloe11} with integration time steps of $dt=10^{-4}$ in units of $1/D_r$. To realize the boundary anchoring on the unit disk, boundary particles did not participate in the stochastic dynamics but kept the fixed director angle profile given in Eq.~(\ref{eq:bcpm}) and acted as neighbors for the director dynamics of bulk particles. 

\subsection{Initial conditions}
Random initial conditions have been used for most of the shown simulation results, except for the time series in Fig.~\ref{fig2}c,d, as well as to generate steady state pattern Fig.~\ref{fig4}a~($m=0$) and the initial state $t=0$ in Fig.~\ref{fig5}~($m=2$). To describe the initial conditions for the latter cases, we denote in the following $(r,\theta)$ and $(r_i,\theta_i)$ as the position of a given field or particle position in cylindrical coordinates. The initial point defect states in Fig.~\ref{fig2}c,d ($t=0$) are respectively given by $\Psi_k=e^{i2\theta}$ and $\alpha_i=(2/3)\theta_i$ ($k=3$ in the particle model), which maps to the $k$-atic phase angles $\phi_{(k)}$ and $\phi_{(k),i}$ as shown in Eqs.~(\ref{eq:anglemap}) and~(\ref{eq:micrangmap}). To generate spiral stationary states that do not form spontaneously, we use $h_m(r,\theta)=8\pi(r-1)+m\theta$, and initialize the order parameter field as $\Psi_k=e^{ih_m(r,\theta)}$ and the particle director field as \hbox{$\alpha_i=[h_m(r_i,\theta_i)/3\mod2\pi$]} ($k=3$ in the particle model). The equilibration of these states gives rise to the patterns shown in Fig.~\ref{fig4}a ($m=0$) and to the initial state $t=0$ in Fig.~\ref{fig5} ($m=2$).

\end{document}